\documentclass[]{aa}  

\usepackage{graphicx}
\usepackage{natbib}

\usepackage{txfonts}
\usepackage{hyperref}

\usepackage{amsmath}
\usepackage{comment}
\usepackage{cleveref}
\usepackage{subfigure}
\usepackage{xcolor}

\usepackage{savesym}
\savesymbol{tablenum}
\usepackage{siunitx}
\restoresymbol{SIX}{tablenum}

\newcommand{\revbf}[1]{{#1}}

\begin{document}

    \title{Three-temperature radiation hydrodynamics with PLUTO}

   \subtitle{Tests and applications to protoplanetary disks}

   \author{Dhruv Muley \inst{1}
        \and Julio David Melon Fuksman \inst{1}
          \and Hubert Klahr \inst{1}
          }

   \institute{Max Planck Institut f\"ur Astronomie, Königstuhl 17, Heidelberg, DE 69117\\
              \email{muley@mpia.de}
             }

   \date{Received 5 June 2023; accepted 3 August 2023}

 
  \abstract{In circumstellar disks around T Tauri stars, visible and near-infrared stellar irradiation is intercepted by dust at the disk's optical surface and reprocessed into thermal infrared; this subsequently undergoes radiative diffusion through the optically thick bulk of the disk. The gas component---overwhelmingly dominant by mass, but contributing little to the opacity---is heated primarily by gas-grain collisions. In hydrodynamical simulations, however, typical models for this heating process (local isothermality, $\beta$-cooling, two-temperature radiation hydrodynamics) incorporate simplifying assumptions that limit their ranges of validity. To build on these methods, we develop a ``three-temperature" numerical scheme, which self-consistently models energy exchange between gas, dust, and radiation, as a part of the PLUTO radiation-hydrodynamics code. With a range of test problems in 0D, 1D, 2D, and 3D, we demonstrate the efficacy of our method, and make the case for its applicability to a wide range of problems in disk physics, including hydrodynamic instabilities and disk-planet interaction.
}


   \maketitle
%

\section{Introduction} \label{sec:intro}

Protoplanetary disks---the rotationally-supported excess from the star-formation process---consist primarily of hydrogen and helium gas, with trace molecular species. Despite its preponderance by mass, this gas component has a low frequency-integrated opacity, making direct radiative heating and cooling inefficient. The disk's energy exchange with the radiation field is instead mediated by the dust component, typically taken to constitute just 1\% of its total mass. \cite{Chiang1997} studied this process semi-analytically, finding that incident stellar radiation---which peaks in the visible and near-infrared for classical T Tauri stars---creates a superheated dust layer at a disk's radial $\tau_r = 1$ surface. This layer reprocesses the radiation into the thermal infrared, half of which dissipates into interstellar space and half of which radiatively diffuses through dust grains in the optically thick disk midplane. The surrounding gas then thermally couples to the dust temperature structure via collisions. In turn, the density structure of the gas, and consequently that of the entrained dust that sets the $\tau_r = 1$ surface, adapt to maintain hydrostatic equilibrium.

By far, the most detailed models of protoplanetary disk thermal structure come from physical-chemical codes \citep[e.g.,][]{Woitke2009,Bruderer2012,Bruderer2013,Calahan2021MAPS}, followed by standalone, multiwavelength radiative transfer codes \citep[e.g.,][]{Dullemond2012,Whitney2013}. However, the computational expense of these numerical algorithms, along with their implicit assumption that disks are static on timescales relevant to thermal physics, makes them intractable to include in dynamical problems. Hydrodynamical simulations must therefore use various parametrizations, of which the ``locally isothermal'' assumption---in which the temperature in each grid cell is fixed to its initial value---has been historically important for its simplicity and ease of interpretation. But more recently, simulations of hydrodynamic instabilities \revbf{and the anticyclonic vortices they generate} \ \citep{Malygin2017,Pierens2018,TarczayNehez2020,Manger2021,HuangYu2022}, gaps and rings \citep{Miranda2019,Ziampras2020}, spiral arms \citep{Miranda2020,Miranda2020b}, and circumplanetary disks \citep{Fung2019} have revealed that the choice of disk thermodynamics prescription has a substantial influence on their morphology and behavior. The quality of disk observations continues to improve \citep[e.g.,][]{Keppler2018,Dong2018,Teague2019,Pinte2022}, and meaningfully interpreting them requires a more accurate, self-consistent treatment of disk heating and cooling in hydrodynamic models.

To this end, numerical radiation hydrodynamics (RHD) is promising. For computational efficiency, these simulations generally use the ``grey approximation'', in which radiation energy densities, fluxes, and material opacities are averaged over frequency. One widely-used method in the study of protoplanetary disks \citep[e.g.,][]{Klahr2006,Kley2009,Bitsch2013a, Flock2013, Binkert2023} is flux-limited diffusion \citep[FLD; ][]{Levermore1981}, in which the radiative flux $\vec{F}_r$ is proportional to the gradient of radiation energy density $E_r$, subject to a limiter that ensures the system satisfies the speed-of-light limit, $|\vec{F}_r| \leq cE_r$. Though FLD works well in optically thick regions, its definition of flux is limited in the thin limit, making it unable to produce anisotropic features in the radiation field such as beams or shadows. The M1 approximation \citep{Levermore1984} handles these issues with a separate evolution equation for flux. M1 can accurately reproduce both the optically thick diffusion and optically thin free-streaming limits, but has limitations in situations where free-streaming beams cross one another.

Differentiating between dust and gas temperatures is a comparatively recent development in RHD simulations. The star-formation simulations of \cite{Bate2015}, as well as the thermochemical disk simulations of \cite{Wang2017}, \cite{Wang2019} and \cite{Hu2023}, use equilibrium prescriptions to compute dust temperature. By contrast, \cite{Pavlyuchenkov2013}, \cite{Pavlyuchenkov2015}, and \cite{Vorobyov2020} use iterative Newton-Raphson schemes to self-consistently compute energy exchange between gas, dust, and radiation, while using FLD for radiation transport.

In what follows, we present our own three-temperature scheme, implemented within the PLUTO hydrodynamics code \citep{Mignone2007}. To compute energy exchange, we use Newton-Raphson iterations (albeit with differences from the aforementioned works), while to transport radiation, we use the M1 scheme already developed for PLUTO  \citep{MelonFuksman2019,MelonFuksman2021} to handle both the optically thick and thin regimes in protoplanetary disks. Section \ref{sec:method} describes our numerical method. Section \ref{ref:tests} details our code tests: the first two (0D matter-radiation coupling and 1D Ensman shock tube) have known reference solutions and demonstrate basic function, while the latter two (2D self-shadowing and 3D disk-planet interaction) illustrate applications to realistic disks. Section \ref{sec:conclusion} provides a summation of our work and discusses possible future areas of research.

\section{Method}\label{sec:method}
\subsection{Basic equations}
We develop a strategy to solve the equations of radiation hydrodynamics, including a secondary dust fluid. In this work, we make the simplifying assumptions that dust grains are kinetically well-coupled to the gas (Stokes number $\mathrm{St} \ll 1$) and that they have negligible inertia (globally constant dust-to-gas ratio $f_d \ll 1$)---in other words, both evolve with the same velocity field, and the dust has no momentum.
\begin{subequations}\label{eq:radhydro}
\begin{equation}
    \frac{\partial \rho}{\partial t} + \nabla \cdot (\rho \vec{v}) = 0
\end{equation}
\begin{equation}
    \frac{\partial (\rho \vec{v})}{\partial t} + \nabla \cdot (\rho \vec{v} \vec{v}) = -\nabla p - \rho \nabla \Phi + \vec{S}_m + \vec{G}
\end{equation}
\begin{equation}
    \frac{\partial E_g}{\partial t} + \nabla \cdot (E_g \vec{v}) = -\nabla \cdot ((p + \rho \Phi) \vec{v}) + S_m + X_{gd} + cG_g  + S^{\rm irr}_g
\end{equation}
\begin{equation}
\label{eq:ed_evo}
    \frac{\partial E_d}{\partial t} + \nabla \cdot (E_d \vec{v}) = -X_{gd} + cG_d + S^{\rm irr}_d
\end{equation}
\begin{equation}
\label{eq:er_evo}
    \frac{\partial E_r}{\partial t} + \hat{c}\nabla \cdot \vec{F}_r = -\hat{c}(G_g + G_d)
\end{equation}
\begin{equation}
\label{eq:flux_evo}
    \frac{\partial \vec{F}_r}{\partial t} + \hat{c}\nabla \cdot \mathbf{P}_r = -\hat{c}\vec{G}
\end{equation}
\end{subequations}
where $\rho, \vec{v}, p$ represent the gas density, pressure, and velocity respectively. $\rho_d$ is the dust density, while $f_d$ is the dust-to-gas ratio, $\Phi$ is the gravitational potential, and $\{E_g, E_d, E_r\}$ are \textit{total} energy densities for gas, dust, and radiation respectively. $c$ is the speed of light, while the $\hat{c}$ term is a ``reduced speed of light'' \citep{Gnedin2001} which enables longer timesteps, but must nevertheless exceed all hydrodynamic velocities relevant to the problem \citep{Skinner2013}.

We assume that the gas follows an ideal equation of state,
\begin{equation}
    p = \rho_g k_B T_g/\mu m_H
\end{equation}
where $k_B$ is the Boltzmann constant, $T_g$ the gas temperature, $\mu$ the mean molecular weight, and $m_H$ the mass of a hydrogen atom. The adiabatic index $\gamma \equiv \partial \ln p/\partial \ln \rho$ is a constant, implying an \textit{internal} \revbf{(thermal)} energy density
\begin{equation}
    \xi_g = p/(\gamma - 1)\,.
\end{equation}
\revbf{Because we do not consider the kinetic energy of dust in this work, the internal energy of dust, $\xi_d$, is equal to its total energy $E_d$. For consistency of notation in what follows, we additionally define a symbol $\xi_r \equiv E_r$.}

Dust and gas exchange \revbf{internal} energy collisionally via the term 
\begin{equation}
    X_{gd} \equiv t_c^{-1}(r_{gd} \xi_d - \xi_g)
\end{equation}
where $t_c$ is the thermal relaxation time
and $r_{gd} = k_B/\mu m_H f_d c_d (\gamma - 1)$ is the ratio of heat capacity per unit volume between gas and dust. $c_d$ is the specific heat capacity of dust. We compute the gas cooling time $t_c$ as a function of the dust-gas stopping time $t_s$, in the Epstein regime \citep{BurkeHollenbach83,Speedie2022}
 
\begin{equation}
\label{eq:tcool}
    t_c = \frac{2/3}{\gamma - 1} f_d^{-1} t_s \eta^{-1}
\end{equation}
where $\eta$ is an order-unity ``accommodation coefficient'' which we henceforth ignore. The corresponding dust cooling time is given by

\begin{equation}
\label{eq:tcool_d}
    t_{c, \rm dust} = r_{\rm gd}^{-1} t_c = \frac{2}{3}\frac{c_d}{k_B/\mu m_H} t_s\,.
\end{equation}

In a realistic disk, the stopping time is computed by averaging over a grain-size distribution, which we describe as part of Section \ref{sec:ssi}. The usual two-temperature radiation hydrodynamics represents the limit where $t_c \rightarrow 0$.

\begin{subequations}\label{eq:rad_interaction}
Net radiative heating/cooling is given for gas by
\begin{equation}
\begin{split}
    G_g \equiv &-\rho \kappa_{g}(a_r T_g^4 - \xi_r) \\
&- \rho (2 \kappa_{g} - \chi_g) \vec{\beta} \cdot \vec{F}_r -\rho \chi_{g} \vec{\beta} \cdot (\xi_r \vec{\beta} + \vec{\beta} \cdot \mathbf{P}_r)
\end{split}
\end{equation}
and likewise for dust by
\begin{equation}
\begin{split}
    G_d \equiv &-\rho \kappa_{d} f_d (a_r T_d^4 - \xi_r) \\ &- \rho f_d (2 \kappa_{d} - \chi_d) \vec{\beta} \cdot \vec{F}_r -\rho f_d \chi_{d} \vec{\beta} \cdot (\xi_r \vec{\beta} + \vec{\beta} \cdot \mathbf{P}_r)
\end{split}
\end{equation}
Similarly, the attenuation of radiative flux by gas is given by
\begin{equation}
\begin{split}
    \vec{G}_g \equiv & \rho \chi_g \vec{F}_r -\rho \kappa_{g}(a_r T_g^4 - \xi_r) \vec{\beta}\\
&- 2 \rho \kappa_{g} (\vec{\beta} \cdot \vec{F}_r) \vec{\beta} -\rho \chi_{g} (\xi_r \vec{\beta} + \vec{\beta} \cdot \mathbf{P}_r)
\end{split}
\end{equation}
and for dust by
\begin{equation}
\begin{split}
    \vec{G}_d \equiv & \rho f_d \chi_d \vec{F}_r -\rho f_d \kappa_{d}(a_r T_d^4 - \xi_r) \vec{\beta}\\
&- 2 \rho f_d \kappa_{d} (\vec{\beta} \cdot \vec{F}_r) \vec{\beta} -\rho f_d \chi_{d} (\xi_r \vec{\beta} + \vec{\beta} \cdot \mathbf{P}_r)
\end{split}
\end{equation}
involving absorption opacities $\kappa$, total (absorption plus scattering) opacities $\chi$, temperatures $T$, the radiation constant $a_r = 4\sigma_{\rm SB}/c$, and the radiation pressure tensor $\mathbf{P}_r$. Typically, $\kappa$ is computed as a Planck average over frequency, while $\chi$ is computed as a Rosseland average. We define $\vec{G} \equiv \vec{G}_g + \vec{G}_d$; because our current implementation assumes dust follows the gas and has negligible inertia, it is this total $\vec{G}$ that gets added to the gas momentum equation and integrated during the hydrodynamic step.

Each of the equations (\ref{eq:rad_interaction}) can be divided into a $\vec{\beta}$-independent and $\vec{\beta}$-dependent part, $G = G^{\rm w} + G'(\vec{\beta})$. The $G'(\vec{\beta})$ parts arise from a Lorentz transformation of the radiation-interaction terms from the fluid's comoving frame to the observer's frame, truncated in the mildly relativistic regime \citep{Skinner2013,MelonFuksman2019}. In protoplanetary disks, $\beta \ll 1$ \footnote{The Keplerian speed of an object orbiting a $1 M_\odot$ at 0.01 au (~1 $R_\odot$ for a T Tauri star)---essentially the maximum attainable in a protoplanetary-disk context---is $\beta \approx 5 \times 10^{-3}$.}, so these terms are small and can be integrated explicitly.

\end{subequations}

The radiation pressure tensor $\mathbf{P}_r(E_r, \vec{F}_r)$ is evaluated according to the M1 closure \citep{Levermore1984}, which accurately reproduces the free-streaming and diffusion limits:
\begin{equation}
    \mathbf{P}_r^{ij} = D^{ij} E_r = E_r\left(\frac{1 - \Xi}{2} \delta^{ij} + \frac{3\Xi - 1}{2} n^i n^j\right)
\end{equation}
in which
\begin{equation}
\label{eq:xi_edd}
    \Xi = \frac{3 + 4w^2}{5 + 2 \sqrt{4 - 3 w^2}}
\end{equation}
where $\vec{n} = \vec{F}_r/||\vec{F}_r||$, $w = ||\vec{F}_r||/E_r$, and $\delta^{ij}$ is the Kronecker delta.

$S^{\rm irr}_d$ and $S^{\rm irr}_g$ represent radiation emitted directly by the star, which has a higher-frequency spectrum than the dust-emitted radiation represented by $E_r$; the source terms are computed (e.g., via ray tracing) at every hydrodynamic timestep, rather than being evolved dynamically as $E_r$ and $F_r$ are. $S_m$ and $\vec{S}_m$ represent, respectively, energy and momentum source terms such as thermal conductivity and viscosity that are already implemented into PLUTO and integrated during the hydrodynamic step.

\subsection{Conservation laws}
With the reduced speed of light approximation, and ignoring non-ideal source terms other than $\vec{G}_g, \vec{G}_d, G_g, G_d, X_{\rm gd}$, our system conserves the following modified total energy and momentum:
\begin{equation}
    E_{\rm tot} = E_r (c/\hat{c}) + E_g + E_d = \xi_r (c/\hat{c}) + \xi_g + \xi_d + \rho \vec{v}^2/2
\end{equation}
\begin{equation}
    \vec{P}_{\rm tot} = \vec{F}_r/\hat{c} + \vec{p}_g
\end{equation}
which reduce to the standard definitions when $c = \hat{c}$. As mentioned earlier, \cite{Skinner2013} verified that this modification does not impact the dynamics of the system, provided that $\hat{c}$ is much larger than hydrodynamic velocities characteristic of the system.
\subsection{Operator splitting and timestepping}
We solve the dusty radiation-hydrodynamic equations (\ref{eq:radhydro}) using a Strang split, sandwiching a full step of the hydrodynamic terms $\Delta t$ between two half-steps $\Delta t/2$ of the radiative terms. In quasi-conservative form:
\begin{subequations}

\begin{equation}
    \frac{\partial \mathcal{U}}{\partial t} + \left(\nabla \cdot \mathcal{F}_{\rm HD} + \nabla \cdot \mathcal{F}_{\rm rad}\right) = \left(\mathcal{S_{\rm HD}} + \mathcal{S_{\rm rad}^{\rm imp}} + \mathcal{S_{\rm rad}^{\rm exp}}\right)
\end{equation}
where $\mathcal{U}$ is a vector of conserved variables, $\mathcal{S}$ are source terms and $\mathcal{F}$ are advective fluxes:
\begin{equation}
    \mathcal{U} = \left(\rho, \rho \vec{v}, E_g, E_d, E_r, \vec{F}_r\right)^\top
\end{equation}
\begin{equation}
    \mathcal{F}_{\rm HD} = \left(\rho\vec{v}, \rho \vec{v} \vec{v} + p \mathbf{I}, (E_g + p + \rho\Phi) \vec{v}, E_d \vec{v}, 0, 0 \right)^\top
\end{equation}
\begin{equation}
    \mathcal{F}_{\rm rad} = \left(0, 0, 0, 0, \hat{c} E_r, \hat{c} \mathbf{P}_r \right)^\top
\end{equation}
\begin{equation}
    \mathcal{S}_{\rm HD} = \left(0, -\rho \nabla \Phi + \vec{S}_m, S_m, 0,0,0\right)^\top
\end{equation}
\begin{equation}
\begin{split}
     \mathcal{S}_{\rm rad}^{\rm imp} = \left(0, \vec{G}^{\rm w}, X_{gd} + cG^{\rm w}_g + S^{\rm irr}_g,\right.\\ 
     \left.-X_{gd} + cG_d + S^{\rm irr}_d,-\hat{c}(G_g + G_d),-\hat{c}\vec{G}\right)^\top   
\end{split}
\end{equation}
\begin{equation}
     \mathcal{S}_{\rm rad}^{\rm exp} = \left(0, \vec{G}', cG_g',cG_d',-\hat{c}(G_g'+G_d'),-\hat{c}\vec{G}'\right)^\top   
\end{equation}
\end{subequations}
The overall timestep duration $\Delta t$ is set by applying the Courant criterion to the hydrodynamical signal speeds $v_{\rm HD}$. But because $\hat{c} \gg v_{\rm HD}$, the radiation half-steps $\Delta t/2$ must be further divided into substeps of duration $\Delta t_{\rm rad}$ to ensure that the radiation subsystem meets its own Courant criterion. The integration of these substeps can take place through one of two implicit-explicit methods, IMEX1:
\begin{equation}\label{eq:imex1}
\begin{split}
    \mathcal{U}^{(1)} &= \mathcal{U}^n + \mathcal{R}_{\rm rad}^n\Delta t_{\rm rad} + \mathcal{S}_{\rm rad}^{\rm imp(1)}\Delta t_{\rm rad}\\
    \mathcal{U}^{(2)} &= \mathcal{U}^{(1)} + \mathcal{R}_{\rm rad}^{(1)}\Delta t_{\rm rad} + \mathcal{S}_{\rm rad}^{\rm imp(2)}\Delta t_{\rm rad}\\
    \mathcal{U}^{n+1} &= \frac{1}{2}\left(\mathcal{U}^{(1)} + \mathcal{U}^{(2)}\right)
\end{split}
\end{equation}
or IMEX-SSP(2,2,2), both analogous to second-order Runge-Kutta \citep{Pareschi2005}:
\begin{equation}\label{eq:imex-ssp222}
\begin{split}
    \mathcal{U}^{(1)} &= \mathcal{U}^{n} + \mathcal{S}_{\rm rad}^{\rm imp(1)}a\Delta t_{\rm rad}\\
    \mathcal{U}^{(2)} &= \mathcal{U}^{n} + \mathcal{R}_{\rm rad}^{\rm imp (1)}\Delta t_{\rm rad}\\ &+ \mathcal{S}_{\rm rad}^{\rm imp(1)}(1-2a)\Delta t_{\rm rad} + \mathcal{S}_{\rm rad}^{\rm imp(2)}a\Delta t_{\rm rad}\\
    \mathcal{U}^{n+1} &= \mathcal{U}^{n} + \frac{\Delta t_{\rm rad}}{2}\left(\mathcal{R}_{\rm rad}^{(1)} + \mathcal{R}_{\rm rad}^{(2)}\right) \\&+ \frac{\Delta t_{\rm rad}}{2}\left(\mathcal{S}_{\rm rad}^{\rm imp(1)} + \mathcal{S}_{\rm rad}^{\rm imp(2)}\right)
\end{split}
\end{equation}
where $\mathcal{R}_{\rm rad}$ is a shorthand for $-\nabla \cdot \mathcal{F}_{\rm rad} + \mathcal{S_{\rm exp}^{\rm imp}}$ and $a = 1 - 1/\sqrt{2}$. $\mathcal{R}_{\rm rad}$ operators are integrated explicitly. However, the $\mathcal{S}^{\rm imp}_{\rm rad}$ operators, incorporating stiff terms such as opacity and gas-grain collision with characteristic timescales potentially much shorter than advective timescales, are integrated implicitly and are in general functions of $t_{\rm rad}$. 

In what follows, we concern ourselves with solving the implicit source terms. The explicit, advective radiative terms $\mathcal{F}_{
\rm rad}$, encompassing the transport of radiation energy density by radiative flux $\hat{c} \vec{F}_r$, and of radiative flux by the radiation pressure tensor $\hat{c} \mathbf{P}_r$, are updated using a Godunov scheme and Riemann solver as described in \cite{MelonFuksman2019}.

\subsection{Momentum source terms}
We update the radiative flux using the first-order implicit step
\begin{equation}
    \vec{F}_r^{i+1} - \vec{F}_r^{i} = -\hat{c}\vec{G}^{i+1}\delta t = - \hat{c} \rho \delta t (\chi_d f_d + \chi_g) \vec{F}_r^{i+1}
\end{equation}
where $\delta t$ equals $\Delta t_{\rm rad}$ times the appropriate order-unity pre-factor for the implicit steps of equations \ref{eq:imex1} and \ref{eq:imex-ssp222}, This can be solved analytically, by rearrangement.
The resulting change in fluid momentum
\begin{equation}
    (\rho \vec{v})^{i+1} - (\rho \vec{v})^{i} = -\hat{c}^{-1}\left(\vec{F}_r^{i+1} - \vec{F}_r^{i}\right) \delta t
\end{equation}
can then be found straightforwardly by conservation.

From the final momenta, we can obtain the change in kinetic energy for gas
\begin{subequations}
\begin{equation}
    K_{g}^{i+1} - K_{g}^{i} = \left[(\vec{p}_{g}^{i+1})^2 - (\vec{p}_{g}^{i})^2\right]/2\rho
\end{equation}
\end{subequations}
in which we make use of the fact that $\rho$ is unchanged during a radiation timestep.

When the total dust and gas opacities $\chi$ are constant, the above equations must only be solved once. In general, however, they can be functions of temperature and density, and so must be solved before every iteration of the energy source terms. We describe these in the following section.

\subsection{Energy source terms}
The exchange of internal energies $\xi$ between dust, gas, and radiation, is updated as follows:
\begin{subequations}
\begin{equation}\label{eq:edeq}
\begin{split}
     \xi_d^{i+1} - \xi_d^{i} = &-(t_c^{-1})^{i + 1} \left[r_{gd} \xi_d^{i+1} - \xi_g^{i+1}\right] \delta t \\ &- (t_d^{i+1})^{-1}\left[a_r (T_d^{i+1})^4 - \xi_r^{i+1} \right]\delta t
\end{split}
\end{equation}
\begin{equation}\label{eq:egeq}
\begin{split}
     \xi_g^{i+1} - \xi_g^{i} = &+(t_c^{-1})^{i + 1} \left[r_{gd} \xi_d^{i+1} - \xi_g^{i+1}\right] \delta t \\ &- (t_g^{i+1})^{-1}\left[a_r (T_g^{i+1})^4 - \xi_r^{i+1}\right]\delta t
\end{split}
\end{equation}
with the dust temperature $T_d = \xi_g/(\rho c_d f_d)$ and the gas temperature $T_g = \xi_g/(\rho k_B/\mu)$. To simplify notation, we have introduced the symbols $t_d = (c\rho\kappa_df_d)$ (the timescale required for light to travel one mean free path through dust) and $t_g = (c\rho\kappa_g)^{-1}$ (the same, for gas); the relevant absorption opacities $\kappa$ and cooling times $t_c$ are in general functions of temperature and density. $\delta t$ indicates the length of an implicit partial step, which is $\Delta t_{\rm rad}$ in the case of IMEX1, (\ref{eq:imex1}) and $a \Delta t_{\rm rad}$ or $(1-2a) \Delta t_{\rm rad}$, depending on the specific partial step, in the case of IMEX-SSP(2,2,2) (\ref{eq:imex-ssp222}).

The $i$ superscripts indicate quantities at the beginning of the implicit partial step, while those with $i+1$ indicate those at the end. $\xi_r^{i+1}$ is not computed directly in the above system, but is given by conservation of energy, modified to account for the reduced speed of light and the irradiation source terms $S^{\rm irr}_g$ and $S^{\rm irr}_d$:
\begin{equation}\label{eq:ereq}
\begin{split}
    \xi_r^{i+1} - \xi_r^i = &-(\hat{c}/c)\left[(\xi_d^{i+1} - \xi_d^{i}) + (\xi_g^{i+1} - \xi_g^{i})\right] \\
    & + (\hat{c}/c)(S^{\rm irr}_g + S^{\rm irr}_d)\delta t
\end{split}
\end{equation}
\end{subequations}
We insert $\xi_r^{i+1}$ from equation \ref{eq:ereq} into equations \cref{eq:edeq,eq:egeq} and solve the system using a multidimensional Newton's method (detailed in Appendix \ref{sec:newtonmethod}). All equations operate on cell-averaged quantities. We note that the values of the stellar irradiation source term $S^{\rm irr}_d$ are updated, by ray-tracing, only at the start of each overall timestep $\Delta t$, even though their contribution to the energy density is added during each radiation substep. For the tests presented here, we set $S^{\rm irr}_g = 0$, but we foresee it being useful in the future to model photoelectric or line heating in the upper disk \citep{Dullemond2007}.

Our implementation has the advantage of conserving total modified energy to machine precision, and converging within a small number of iterations (typically $\lesssim 5-10$ in disk-planet interaction simulations, even with temperature- and density-dependent cooling times). Moreover, it accommodates temperature change that is rapid compared to  $\Delta t_{\rm rad}$, making it useful in nonequilibrium cases such as accretion shocks, subject to constraints set by the reduced speed of light approximation.

\section{Tests}\label{ref:tests}
To verify the accuracy of our scheme and demonstrate its applicability to protoplanetary disks, we run test problems in 0D, 1D, 2D, and 3D. For the tests we present, we use third-order Runge-Kutta (RK3) timestepping with third-order weighted essentially non-oscillatory (WENO3) reconstruction \citep{Yamaleev2009} and a Harten-Lax-van Leer-Contact (HLLC) Riemann solver for both the gas and radiation. The sole exception is our 2D disk self-shadowing test, in which large contrasts in temperature, density, and velocity emerge in high, close-in regions of the disk. As such, for this problem, we opt for second-order Runge-Kutta (RK2) time integration and piecewise linear method (PLM) spatial reconstruction with a van Leer flux limiter.

\subsection{0D test: radiation-matter coupling}
\begin{figure}
    \centering
    \includegraphics[width=0.999\linewidth]{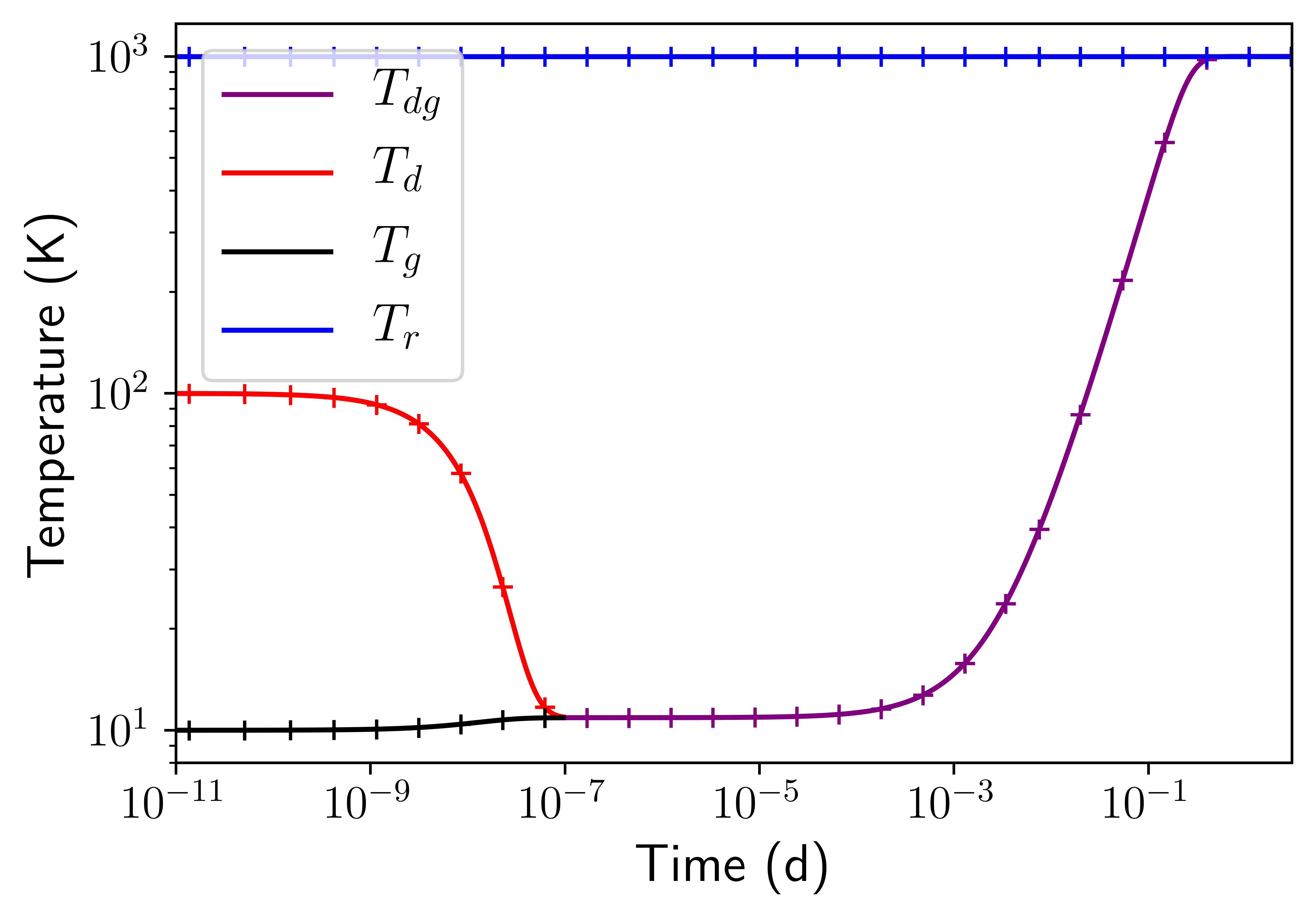}
    \caption{A 0D test of our three-temperature scheme, for which we set the dust-gas stopping time $t_s = 10^{-9} t_0 = 8 \times 10^{-9} \ {\rm d}$ and the opacity $\kappa_d = 3.9 \ \si{\square\cm\per\gram}$. Pluses indicate numerical results, while solid lines indicate analytical solutions taken from equations \ref{eq:col} \textit{(red, black)} and \ref{eq:absemcol} \textit{(purple)}. Because the system is dominated by radiation energy, the radiation temperature \textit{(blue)} is nearly constant throughout the system's evolution.}
    \label{fig:three_temp_0d}
\end{figure}

For our first set of tests, we consider a zero-dimensional setup probing only the local evolution of gas, dust, and radiation temperatures. We turn off the hydrodynamic equations and radiation advective terms, and set irradiation source terms and radiative flux to zero, in order to isolate the effects of radiation-matter coupling. In this case, we can write the evolution of internal energies concisely in terms of $\vec{\xi} = (\xi_d, \xi_g)^\top$:
\begin{equation}\label{eq:inten_matrix}
    \frac{\partial \vec{\xi}}{\partial t} = \left[\mathbf{C}\vec{\xi}\right] + \left[\vec{A}_{\rm tot} + \mathbf{A}\vec{\xi} +  \vec{Q}(\vec{\xi})\right]
\end{equation}
where $\mathbf{A}$ includes linear absorption terms, $\mathbf{C}$ the collisional terms, $\vec{A}_{\rm tot}$ the constant absorption terms, and $\vec{Q}$ the nonlinear emission terms:
\begin{subequations}
\begin{equation}
    \mathbf{C} = \begin{bmatrix}
-t_c^{-1} r_{gd} & +t_c^{-1}\\
+t_c^{-1} r_{gd} & -t_c^{-1}
\end{bmatrix}
\end{equation}
\begin{equation}
    \mathbf{A} = \frac{\hat{c}}{c}\begin{bmatrix}
-t_d^{-1} & -t_d^{-1}\\
-t_g^{-1} & -t_g^{-1}
\end{bmatrix}
\end{equation}
\begin{equation}
    \vec{A}_{\rm tot} = \xi_{\rm tot}\hat{c}/c\left(t_d^{-1}, t_g^{-1}\right)^\top
\end{equation}
\begin{equation}
    \vec{Q}(\vec{\xi}) = -a_r \left(t_d^{-1} T_d^4, t_g^{-1} T_g^4\right)^\top
\end{equation}
\end{subequations}

For our setup, we use a gas density of $\rho_g = 7.78 \times 10^{-18} \si{\gram\per\cubic\cm}$ with a dust-to-gas ratio $f_d = 10^{-2}$ and stopping time $t_s = 10^{-9} t_0$,  where $t_0 = 8 \ {\rm days}$ is the code unit for time. We use a fixed opacity $\kappa_d = 3.9 \ \si{\square\cm\per\gram}$ and $\hat{c} = c$. At $t = 0$ we set $T_g = 10^1 \ \si{\kelvin}$, $T_g = 10^2 \ \si{\kelvin}$, and $T_r = 10^3 \ \si{\kelvin}$. The specific heat capacity of the gas $c_g$ is defined by its mean molecular weight $\mu = 1$ and adiabatic index $\gamma = 7/5$; we set the dust specific heat capacity $c_d$ equal to that of the gas. \revbf{By analogy with the determination of Stokes number ${\rm St} \equiv t_s \Omega$ of quasi-Keplerian disks, we can compute a characteristic timescale for this system, $t_{\rm cross} = l_{\rm cross}/v_{\rm cross}$, where $l_{\rm cross}$ and $v_{\rm cross}$ are the length and velocity scales of interest. In this case, we choose a $l_{\rm cross} = (\kappa_d f_d \rho_g)^{-1} = 3.38 \times 10^{18} \ \si{\cm}$---the mean free path of the system---and a $v_{\rm cross} = (k_B T_g/\mu m_p)^{1/2} = 287 \ \si{\meter\per\second}$. This yields a $t_{\rm cross} = 3.6 \times 10^6 \ {\si y}$, or a $t_s/t_{\rm cross} = 5.87 \times 10^{-18}$.}

The system should relax to an equilibrium state in which dust, gas, and radiation have equivalent effective temperatures---specifically, $\xi_d = \xi_g/r_{gd} = c_g \rho_g (\xi_r/a_r)^{1/4}$. Furthermore, in several special cases, it has analytical solutions against which our numerical scheme can be tested. For instance, when collisional energy exchange between gas and dust happens much faster than absorption, the system has a solution in terms of matrix exponentials:
\begin{equation} \label{eq:col}
    \vec{\xi}(t) \approx \exp \left[\mathbf{C}t\right]\vec{\xi}_0
\end{equation}
Given our simulation parameters, we find that equation \ref{eq:col} is a good description of the temperature evolution at early times. We plot its predictions for $T_g$ and $T_d$ as black and red lines respectively in \ref{fig:three_temp_0d}. On timescales $t > t_c/(1 + r_{gd}) \approx t_s$, such a system reaches a quasi-steady state where $\xi_d = \xi_g/r_{gd}$---in other words, the temperatures of dust and gas become equal. 

By contrast, when the absorption and emission terms dominate over collisions, and additionally the system is radiation-energy dominated throughout its evolution (i.e., $(c/\hat{c}) \xi_r \equiv (\xi_{\rm tot} - \xi_d - \xi_g) \gg \xi_g + \xi_d$), $\hat{c}$ drops out of the equations and we get:
\begin{equation}\label{eq:absem}
\frac{\partial}{\partial t} \begin{bmatrix} \xi_d
 \\ \xi_g
\end{bmatrix} \approx \begin{bmatrix} t_d^{-1}(\xi_{r,0} - a_r(\xi_d/\rho f_d c_d)^4)
 \\ t_g^{-1}(\xi_{r,0} - a_r(\xi_g/\rho c_g)^4)
\end{bmatrix}
\end{equation}
Each element of \ref{eq:absem} is identical to the form found in eq. (69) and Figure 5 of \cite{MelonFuksman2019}, and yields an implicit analytic relation that can be inverted by root-finding:
\begin{equation}\label{eq:impsol_rad}
\begin{split}
    2Mt &= \textup{Re } (\arctan(\omega) + \textup{arctanh}(\omega))\\ &- \textup{Re } (\arctan(\omega_0) + \textup{arctanh}(\omega_0))
\end{split}
\end{equation}
where $M_d = (a_r/\xi_{r,0})^{1/4} \xi_{\rm r, 0} t_d^{-1}/(\rho f_d c_d)$, $\omega_d = (a_r/\xi_{r,0})^{1/4}(\xi_d/f_d \rho c_d)$, and $\omega_{d,0} = \omega_d(t = 0)$; $M_g$ and $\omega_g$ are obtained likewise. $\xi_r$ is assumed constant throughout the evolution of the system.

When the collision timescale is much shorter than the absorption-emission timescale, but the dust and gas are already in collisional equilibrium, we can obtain yet another special case by summing the two rows of \cref{eq:absem}:
\begin{equation}\label{eq:absemcol}
    \frac{\partial \xi_d}{\partial t} = \frac{t_d^{-1} + t_g^{-1}}{1 + r_{\rm gd}}(\xi_{r,0} - a_r(\xi_d/\rho c_d f_d)^4)
\end{equation}
where we constrain $\xi_d = \xi_g/r_{gd}$. The solution to equation (\ref{eq:absemcol}) is also in the form of (\ref{eq:impsol_rad}), with $M_{dg} = (a_r/\xi_{r,0})^{1/4} \xi_{r,0} (t_d^{-1} + t_g^{-1})((1 + r_{gd})\rho f_d c_d)^{-1}$ and $\omega_{dg} = (a_r/\xi_{r,0})^{1/4}(\xi_d/\rho f_d c_d)$.

We plot equation (\ref{eq:absemcol}) as a purple curve in Figure \ref{fig:three_temp_0d}. Our scheme accurately handles the regime where dust and gas temperatures are well-coupled, showing that it is effective at reproducing the ``two-temperature'' limit of typical radiation-hydrodynamics simulations.

\subsection{1D test: Subcritical dusty radiative shock}

Having verified that our scheme accurately reproduces three-temperature static thermal coupling, we seek to study its performance in a a dynamical setting. To this end, we run a modified version of the 1D subcritical radiative shock in \cite{Ensman1994}, a standard test for radiation-hydrodynamic codes 
\citep[see e.g., section 4.1 of][and references therein]{MelonFuksman2021} incorporating sharp transitions in temperature and opacity.

In this setup, our domain spans the interval $[0, 7 \times 10^{10} \ \si{\cm}]$. The initial gas density is $\rho_g = 7.78 \times 10^{-10} \si{\gram\per\cubic\cm}$, with a dust-to-gas ratio $f_d = 10^{-2}$. Dust, gas, and radiation temperatures are all initialized in equilibrium, $T_r = T_g = T_d = 10 \ \si{\kelvin}$; $\gamma = 7/5$, $\mu = 1$, and $c_d = c_g$ are the same as in the 0D case. The dust absorption opacity is set such that the inverse mean free path of radiation through dust, $l_d^{-1} \equiv \kappa_d \rho_d = 3.1 \times 10^{-10} \si{\per\cm}$, with dust scattering opacity $\chi_d - \kappa_d$, and gas total opacity $\chi_g$, both assumed to be zero. To form the shock, given our otherwise equilibrium initial conditions, we set an initial gas velocity $v_x = -6 \ \si{\km\per\second}$, and impose a reflective condition on the left boundary. We fiducially choose a reduced speed of light $\hat{c} = 10^{-3} c$ and a spatial resolution $N_x = 1200$ cells; these figures give results consistent with those of our tests (not shown) at $\hat{c} = c$, and both higher and lower resolutions.

We test stopping times $t_s = \{10^{-7}, 10^{-4}, 10^{-1}\} t_0$, where we again use $t_0 = 8 \ {\rm d}$ as the code unit for time. \revbf{For this problem, we define the characteristic timescale using $l_d$ as our length scale and $v_x$ as the velocity scale, yielding a $t_{\rm cross} = 7.78 \times 10^{-2} t_0$ and $t_s/t_{\rm cross} = 1.28 \times \{10^{-5}, 10^{-2}, 10^{1}\}$. The last case is included to demonstrate the properties of our code when dust and gas have weak \textit{thermal} coupling, although the long $t_s/t_{\rm cross}$ would also allow for relative motion between dust and gas, which we do not model.}

\begin{figure}
    \centering
    \includegraphics[width = 0.999\linewidth]{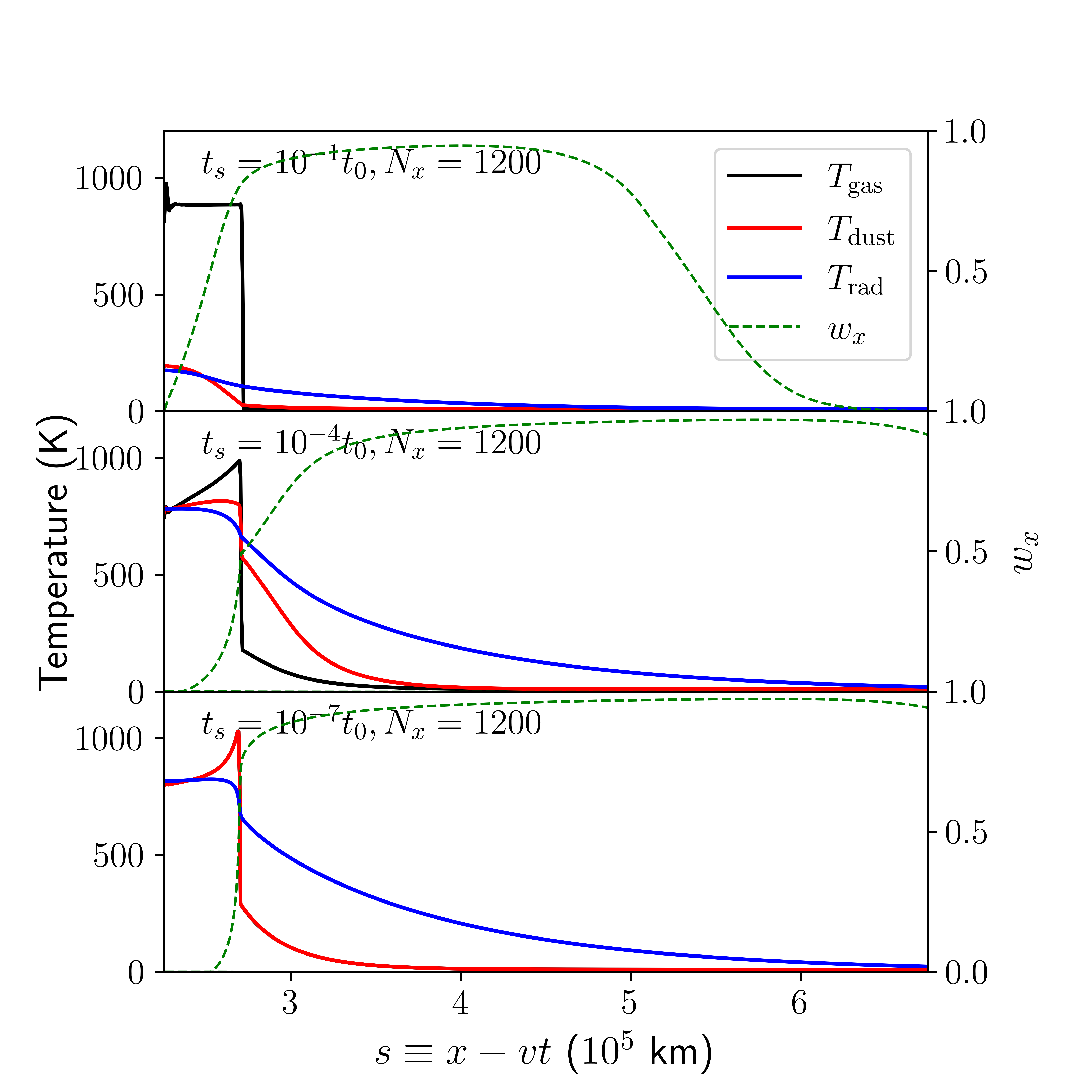}
    \caption{Temperature profiles in the dusty radiative shock test at $t = 0.054 t_0 = 3.75 \times 10^4 \ \si{\sec}$, for a variety of stopping times. For $t_s = 10^{-1}$, the gas temperature is that of a standard hydrodynamic shock, with only weak losses to dust and radiation; for $t_s = 10^{-7}$, $T_d$ and $T_g$ have come into equilibrium, and they (along with $w_x$ and $T_r$) match the results of the two-temperature radiative shock tests in section 4.1 of \cite[][]{MelonFuksman2021} (note that the signed ``reduced flux'' $w_x \equiv F_x/E_r$ is denoted as $f$ in their work).}
    \label{fig:1d_shock_tc}
\end{figure}

In Figure \ref{fig:1d_shock_tc}, we plot temperature snapshots at $t = 0.054 t_0 = 3.75 \times 10^4 \ \si{\sec}$ for all stopping times, selecting $N_x = 1200$ as our fiducial resolution. For $t_s = 10^{-1}$, the gas temperature is that of a standard hydrodynamic shock. In the pre-shock region dust and gas are essentially in equilibrium; in the post-shock region, by contrast, the dust heats at a rate $\dot{T}_{d, \rm coll} \approx (T_g - T_d) t_{\rm c, dust}^{-1} \approx (T_{g,+} - T_{d,-}) t_{\rm c, dust}^{-1}$, where minuses and pluses denote pre- and post-shock quantities respectively. Deeper in the post-shock region, this trend breaks down, as radiative cooling (which scales as $\dot{T}_{d, \rm rad}  \approx -c l_d^{-1} a_r (T_{d,-}^4 - T_{r,-}^4)/c_d \rho_d \ $) becomes sufficient to counteract the effects of collisional heating.

In the $t_s = 10^{-4} t_0$ case, the cooling time $t_{\rm c, dust}$ is short, and $T_d$ increases rapidly immediately after encountering the shock. Because the collisional heating rate is faster than for $t_s = 10^{-1} t_0$, the radiative cooling rate, and thus the maximum post-shock dust temperature, must be correspondingly higher. In the pre-shock region, this process is inverted: radiation propagating from the shock front heats the dust above its initial preshock value, and via collisions, the gas. For the strongly coupled case, with $t_s = 10^{-7} t_0$, dust and gas are in collisional equilibrium, and we recover the two-temperature results of \cite{MelonFuksman2021}. Because the Courant-limited radiation timestep $\Delta t_{\rm rad} \approx \Delta x/\hat{c}$ is much longer than our dust coupling time, this result demonstrates the efficacy of our IMEX scheme in the stiff regime.

In these 1D simulations, changes in the radiation field are driven by collisional heating of the dust, and ultimately by the rightward motion of the shock front---a process much slower than radiation propagation or absorption/emission. In this regime, we can drop the time derivative terms in Equation \ref{eq:flux_evo} and take the nonrelativistic limit

\begin{equation}
    \nabla (\Xi E_r) \approx -\rho_d \kappa_d F_x = -l_d^{-1} F_x
\end{equation}

where the $\Xi$ parameter is defined in Equation \ref{eq:xi_edd} as a function of the ``reduced flux'' $w$. Adding a sign to this reduced flux $w_x = F_x/E_r = \pm w$ and rearranging terms, we get that

\begin{equation}
\label{eq:flux_driving}
     \partial_{w_x} (\Xi) \nabla(w_x) \approx -(l_d^{-1} w_x + l_r^{-1} \Xi)
\end{equation}

where $l_r^{-1} \equiv \nabla (E_r)E_r^{-1}$ is the relative rate of change in the radiation field. In other words, variations in the radiation field on length scales shorter (longer) than some critical scale $l_d \Xi |w_x|^{-1}$ drive (damp) the flux. 

This is clearly visible in Figure \ref{fig:1d_shock_tc}: in the $t_s = 10^{-1} t_0$ case, the relatively moderate gradient in radiation energy means that the transition from the $w_x \approx 0$ diffusion regime to the $w_x \approx 1$ free-streaming regime occurs smoothly over the post-shocked region. For $s \gtrsim 5 \times 10^{5} \ \si{\km}$, radiation from the shock front is sufficiently extincted that the radiation field is dominated by the flat 10 K background; accordingly, $|l_r| \rightarrow \infty $ and $w_x$ damps to zero.

By contrast, the $t_s = 10^{-7} t_0$ case has a flat radiation field for much of the post-shock region, but a sharper one in the immediate vicinity of the shock, creating an abrupt transition between the diffusion and free-streaming regimes. Because the post-shock radiation field is stronger in this case, it propagates to a greater optical depth before decaying into the flat $T_r = 10 {\rm \ K}$ background, so the $w_x$ damping seen in the $t_s = 10^{-1} t_0$ simulation is not visible in the plotted region. The $t_s = 10^{-4} t_0$ case represents an intermediate case between these two extremes.

\subsection{2D test: self-shadowing instability}\label{sec:ssi}
\begin{figure*}
    \centering
    \includegraphics[width=0.999\textwidth]{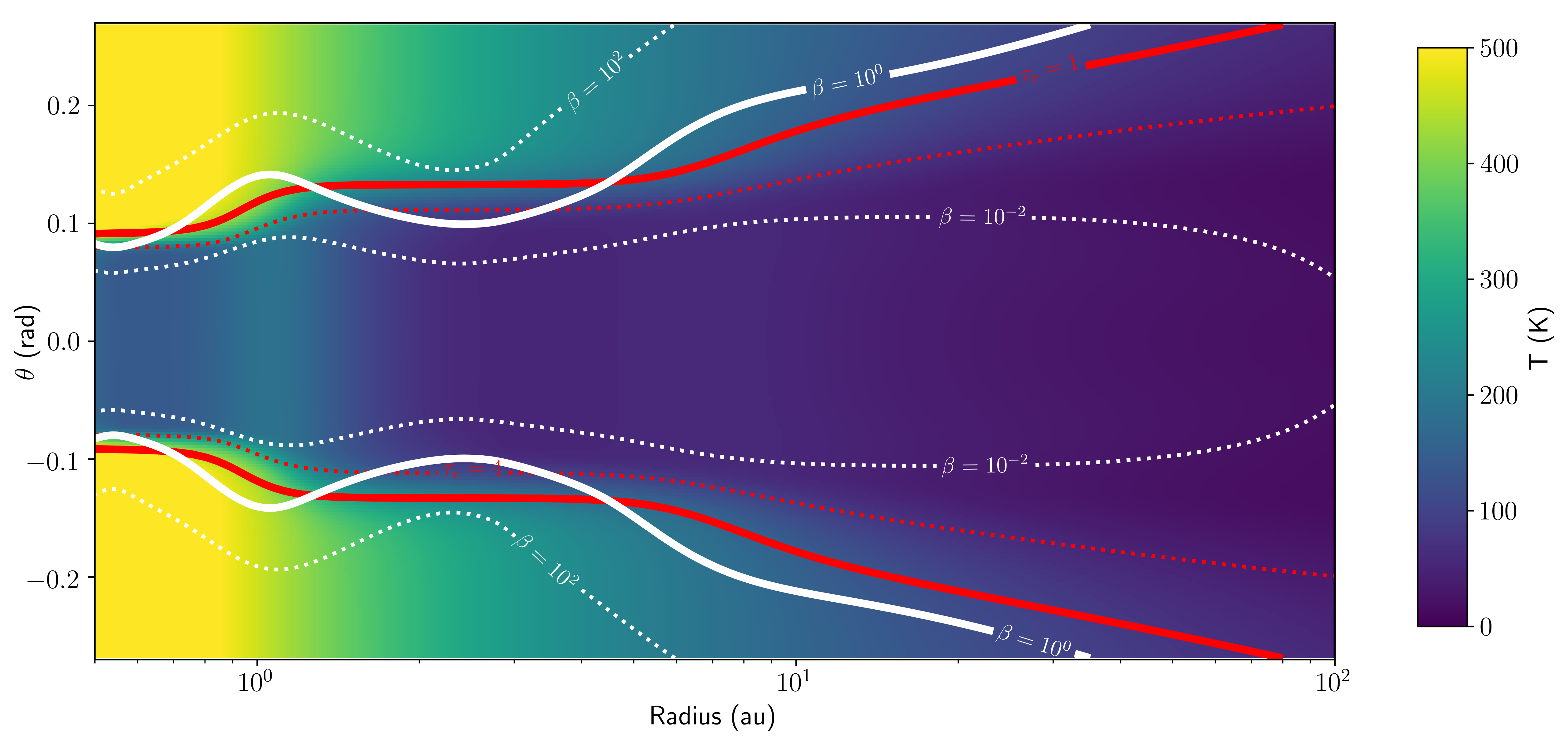}
    \includegraphics[width=0.999\textwidth]{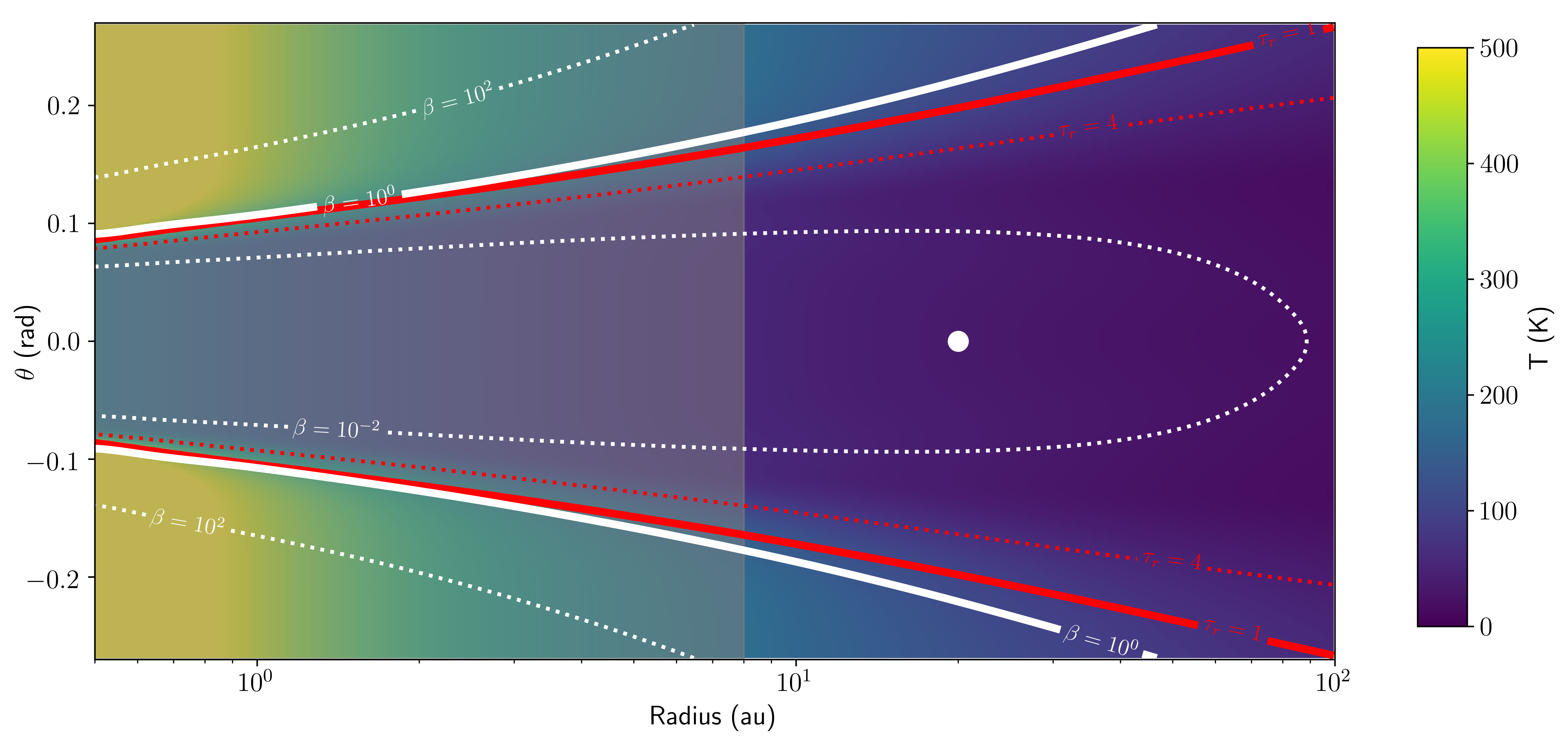}
    \caption{\textit{Above}, initial $T_d = T_g$ for our 2D self-shadowing test (\ref{sec:ssi}), based on \texttt{dg3t100} from \citep{MelonFuksman2022}; \textit{below}, the same for our 3D planet-driven spiral test (\ref{sec:spiral}), based on \texttt{dg3t0.1}. We overplot contours for radial optical depth $\tau_r$ \textit{(red)} (computed using opacities Planck-averaged over the stellar spectrum) and cooling parameter $\beta \equiv t_c \Omega$. Our calculations of these quantities assume an MRN distribution of small silicon-graphite grains, and a $1 M_\oplus$ classical T Tauri star as the source of light and gravity. In the lower plot, a white dot indicates the position of the embedded $1 M_J$ planet, while grey shading covers the region outside the $r = [8, 100]$ au boundaries of our 3D simulation.}
    \label{fig:ssi_ic}
\end{figure*}

\begin{figure*}
    \centering
    \includegraphics[width=0.999\textwidth]{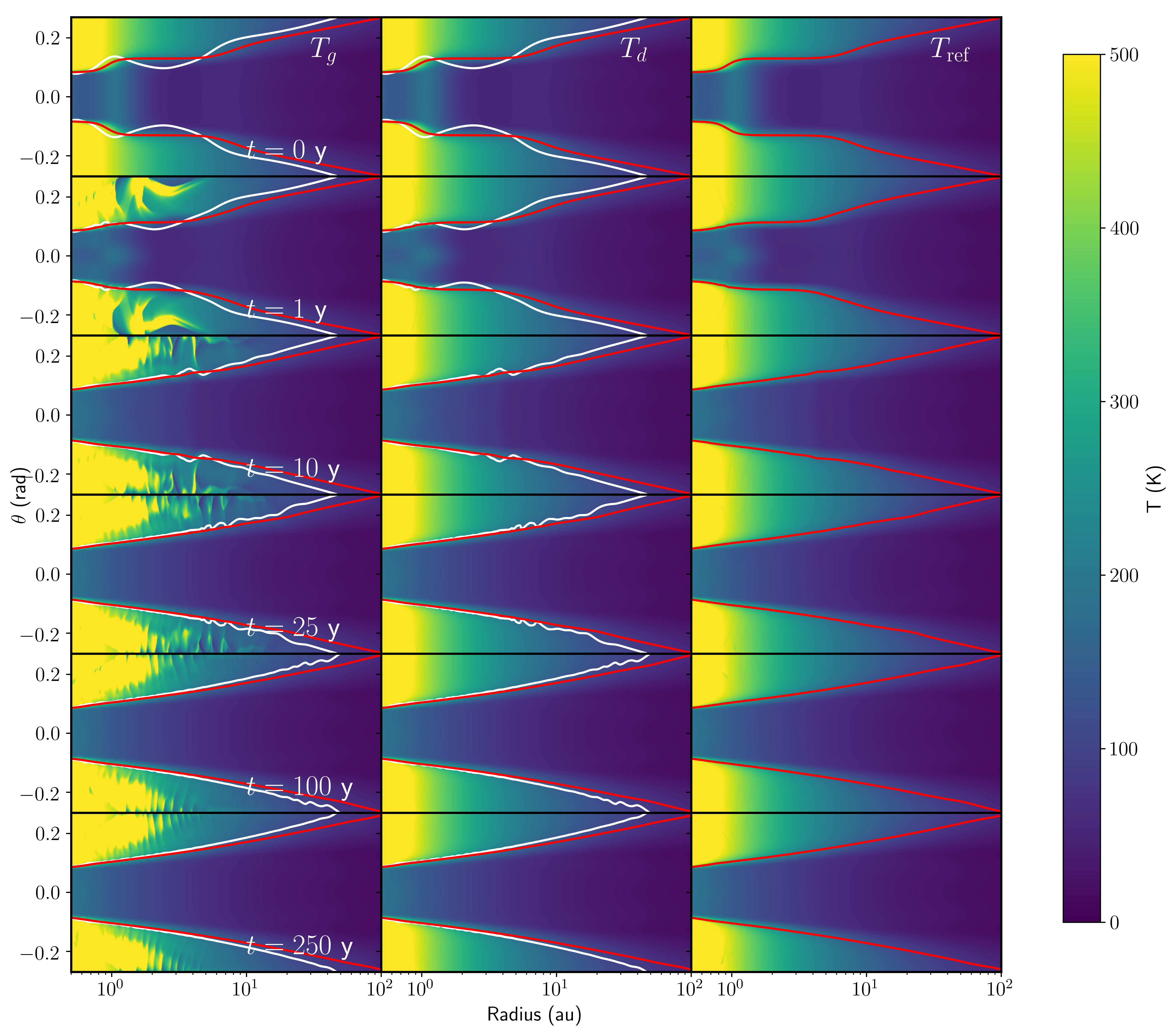}
    \caption{Two-dimensional temperature profiles from our SSI tests. The $T_g$ \textit{(left)} and $T_d$ \textit{(middle)} columns come from our simulations with realistic $t_s$, whereas $T_{\rm ref}$ \textit{(right)} is from our reference simulations with $t_s = (2\pi)^{-1} \times 10^{-10} \ {\rm y} $. For ease of interpretation, we overplot $\beta = 1$ \textit{(white)} and $\tau =1$ \textit{(red)} surfaces. The reference disk relaxes rapidly to a smooth hydrostatic configuration; the bulk of the realistic disk does so as well, although differences persist in the slowly-cooling disk atmosphere.}
    \label{fig:test_self_shadow}
\end{figure*}

The self-shadowing or thermal wave instability \citep[SSI; see e.g.,][]{DAlessio1999,Dullemond2000,Watanabe2008,Flock2020,Wu2021} has been proposed as a planet-free means of generating gaps and rings in circumstellar disks. SSI begins with a disk in initial radiative and (vertical) hydrostatic equilibrium. A small perturbation to the disk's radial $\tau_r = 1$ surface would create slightly raised regions which absorb more stellar irradiation, and lowered, obscured regions which absorb less. To maintain hydrostatic equilibrium, the raised regions would increase in scale height (and thus absorb even more starlight), while the lowered ones would decrease in scale height and absorb even less. 

SSI is readily reproduced in 1+1D simulations that parametrize the vertical extent of a disk \citep[e.g.,][]{Ueda2021}. It can also be generated \citep[e.g.,][]{Ueda2019} by an iterative procedure that first uses radiative transfer to obtain a disk's temperature profile, given its density structure; assuming hydrostatic equilibrium, this profile is then used to recompute the density structure for the next iteration. 

However, \cite{MelonFuksman2022} find that this picture breaks down with additional physics: in their two-temperature radiation hydrodynamics simulations, SSI bumps do not grow from noise, and decay rapidly even when artificially introduced via the aforementioned iterative procedure. They explain their results by arguing that the erosion of perturbations at the $\tau_r = 1$ surface---caused by in-plane radiation transport and the hydrostatic adaptation of the gas in the upper disk---occurs on timescales much faster than the classical vertical radiative heating timescale of the disk \citep{Hubeny1990, Zhu2015, Zhang2020}
\begin{equation}\label{eq:classical_rad_cool}
    t_{\rm SSI, rad} = \frac{3(\tau_v + \tau_v^{-1}) (\Sigma_d c_d + \Sigma_g c_g)}{4c a_r (T^2_{\rm irr} + T^2_{\rm d, mid})(T_{\rm irr} + T_{\rm d, mid})}
\end{equation}
where $\tau_v = \kappa_d (\Sigma_d / 2)$ is the \textit{vertical} optical depth of the disk to the midplane in thermal infrared, $\Sigma_d$ and $\Sigma_g$ are the dust and gas surface densities, $T_{\rm irr}$ is the equilibrium temperature of the stellar irradiation field in the optically thin region of the disk, and $T_{\rm d, mid}$ is the dust midplane temperature. Physically, this means that bumps at the optical surface decay long before they can impact the temperature at the disk surface.

A nonzero coupling time between gas and dust temperatures would allow a temperature perturbation at the $\tau_r = 1$ surface to establish itself in the underlying disk column by propagating only through the dust---a much faster process than that implied by equation (\ref{eq:classical_rad_cool}), in which the gas and dust are heated together. The subsequent adaptation of the gas's vertical structure to this already-established background would conform better to the assumptions of iterative methods which have been successful in reproducing the SSI. Moreover, such a coupling time would delay the response of the perturbed $\tau_r = 1$ gas to in-plane radiation transport, limiting the radial diffusion of SSI bumps. \textbf{We emphasize that nothing about this prediction accounts for the effects of nonzero coupling times between dust and gas \textit{velocities} on the evolution of the SSI, a topic which we defer to future work.}

In the context of protoplanetary disks, a realistic coupling time may be found by computing $t_s$ with the Epstein drag law \citep[see e.g.,][]{Speedie2022,MelonFuksman2023}

\begin{equation}\label{eq:epstein}
    t_s = \frac{\rho_{\rm dg} \left<a_{\rm dg}^3\right>}{\rho_g v_{\rm th}\left<a_{\rm dg}^2\right>}
\end{equation}
where $\rho_{\rm dg}$ is the bulk density of each dust grain and $v_{\rm th} = \sqrt{8\pi k_B T_g/\mu m_H}$ is the gas thermal velocity. A quantity in angle brackets $\left<g(a_{\rm dg})\right>$ represents the integral of that quantity over the (normalized) grain size distribution $n(a)$:
\begin{equation}
    \left<g(a_{\rm dg})\right> = \int_{a_{\rm min}}^{a_{\rm max}} g(a') n(a') da'
\end{equation}

For our three-temperature tests of the SSI, we use the setup \texttt{dg3t100} from \cite{MelonFuksman2022}, generated with the aforementioned iterative procedure, as the basis for our initial conditions. We employ a reduced speed of light $\hat{c} = 10^{-3} c$; our tests at $\hat{c}' = 5 \hat{c} = 5 \times 10^{-3} c$ (not shown) yield converged results. The gas follows an ideal equation of state---with an adiabatic index $\gamma =1.41$ and a mean molecular weight $\mu = 2.3$. We assume a total dust fraction of 1\%, of which 10\% by mass ($f_d = 10^{-3}$) are taken to be small grains relevant to the gas's thermal evolution. These grains are taken to consist of 62.5\% silicate and 37.5\% graphite, with a bulk density $\rho_{\rm gr} = 2.5 \ \si{\gram\per\cubic\cm}$ and a specific heat capacity $c_d = 0.7 \ \si{\joule\per\gram\per\kelvin}$; their sizes follow a classical MRN \citep{MRN77} distribution, $n(a) \propto a^{-3.5}$, with $a_{\rm min} = 5 \ \si{\nano\meter}$ and $a_{\rm max} = 250 \ \si{\nano\meter}$. \footnote{With the MRN distribution, $\left<a_{\rm dg}^3\right>/\left<a_{\rm dg}^2\right> = \sqrt{a_{\rm min} a_{\rm max}} = 35.4 \ \si{\nano\meter}$, and dust fraction $f_{d} \propto \sqrt{a_{\rm max}/a_{\rm min}} - 1$. With these relations, it can be shown that the gas cooling time $t_c$ (equation \ref{eq:tcool}) converges to a fixed limiting value as long as $a_{\rm max} \gg a_{\rm min}$. Physically, this means that collisional energy exchange is dominated by the smallest grains.}

From these grain properties, we compute a temperature-dependent, frequency-integrated dust opacity $\kappa_D(T_d)$, based on the tabulated frequency-dependent figures by \cite{Krieger2020,Krieger2022} as well as $t_s$. We additionally bracket the dust stopping time to lie within $[10^{-10}/2\pi, 10^{1}/2\pi]$ years, and take $T_g \rightarrow \max (T_g, 10 \ \si{\kelvin})$ for the purpose of computing $v_{\rm th}$. We assume a classical T Tauri star with $M_* = 1 M_\odot$, $R_* = 2.0865 R_\odot$, and $T_* = 4000 \ \si{\kelvin}$, to compute the dust irradiation source term $S_d$ as well as the gravitational potential
\begin{equation}
    \Phi_* = -\frac{GM_*}{r_*}
\end{equation}
where $r_* \equiv || \vec{x} - \vec{x} ||$ is the distance between the star and a given point in the domain. We plot the resulting initial condition, as well as relevant contours for radial optical depth $\tau_r$ and gas cooling parameter $\beta \equiv t_c \Omega = (2/3)(\gamma - 1)^{-1} f_d^{-1} t_s \Omega$, in the upper panel of Figure \ref{fig:ssi_ic}. We emphasize that our $\beta$-cooling parameter reflects only gas-grain collisions; because we self-consistently model radiative transfer, we do not incorporate assumptions about time or length scales for radiative diffusion or optically thin cooling, as in some previous work \citep[e.g.,][]{Bae21}.

We run our SSI simulation in 2D axisymmetric polar coordinates, with radial limits at $r = [0.4, 100] \ {\rm au}$ and polar limits at $\theta = \pi/2 + \left[-0.23, 0.23\right]$. This domain is spanned by $\{N_r, N_\theta\} = \{512, 170\}$ grid cells, logarithmically spaced in the radial direction and uniformly spaced in polar angle. At the boundaries, all fields are set equal to those in \texttt{dg3t0.1}, the fiducial smooth hydrostatic setup in \cite{MelonFuksman2022}. We also include a damping zone in the region $\{r < r_{\rm damp, in} \} \cup \{p < 0\}$, where $r_{\rm damp, in} = 0.5 {\rm \ au}$ and $p \equiv -|\theta - \pi/2| + 0.12 + 0.13 (r/{\rm au}) $. In this zone, we enforce $T_d = T_g$, while relaxing $\rho$ and $\vec{v}$ to their smooth hydrostatic values on a timescale
\begin{subequations}
\begin{equation}
    t_{\rm damp} = 0.1 \Omega(r)^{-1} \cos^2\left[\pi/2(1 + p/p_{\rm +})\right]
\end{equation}
in the region where $s < 0$, and 
\begin{equation}
    t_{\rm damp} = 0.1\Omega(r)^{-1} \cos^{-2}\left[\pi/2\left(\frac{r - r_{\rm in}}{r_{\rm damp, in} - r_{\rm in}} \right)\right]
\end{equation}
\end{subequations}
elsewhere in the damping zone; we have tested alternative damping prescriptions and found no substantive change to our results. For comparison with the two-temperature regime, we also run a simulation with a fixed global stopping time $t_{\rm s} = (2\pi)^{-1} \times 10^{-10} {\rm \ y}$, but otherwise identical.

\begin{figure}
    \centering
    \includegraphics[width=0.999\linewidth]{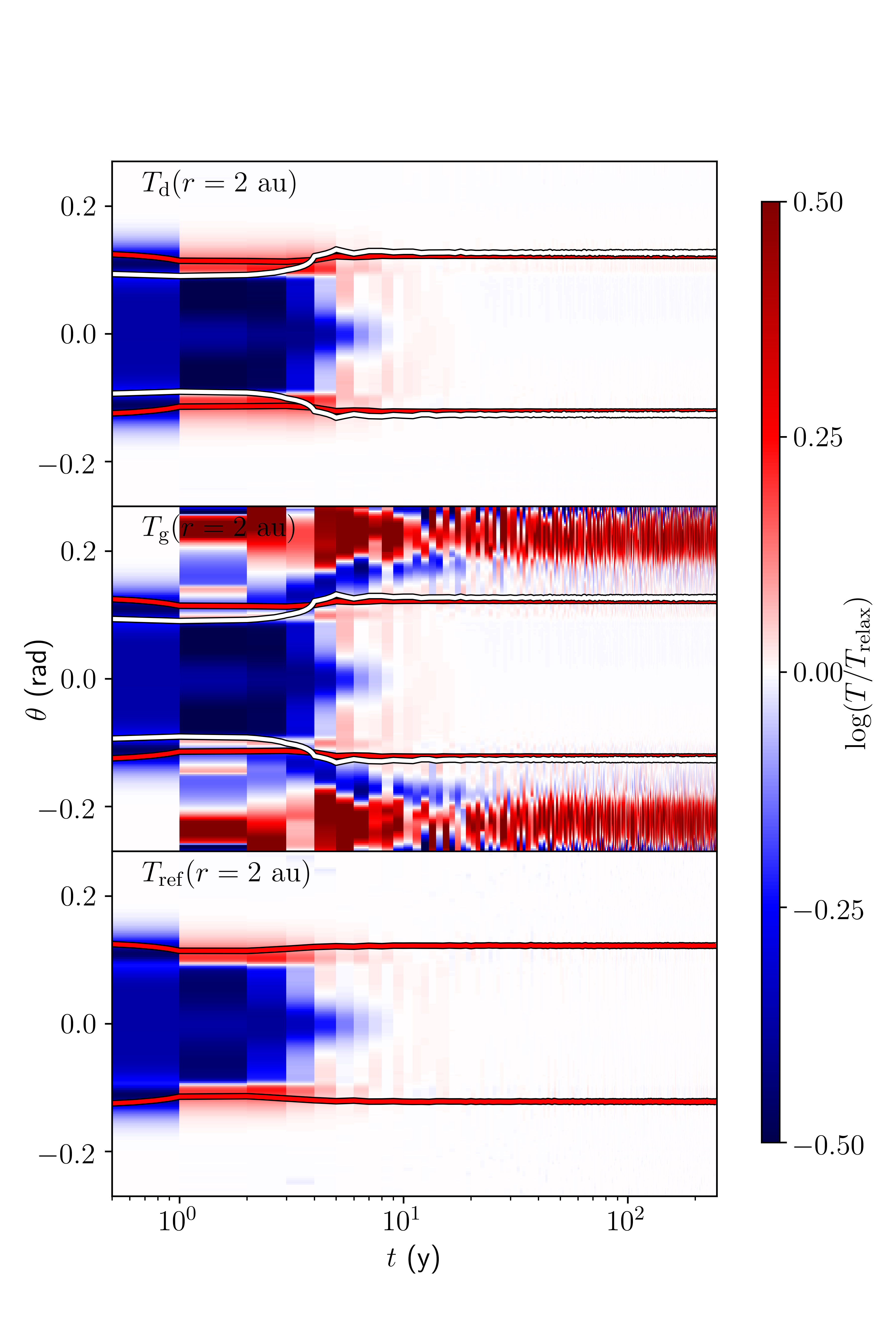}
    \caption{Time evolution of $T_d$ \textit{(top)}, $T_g$ \textit{(center)}, and $T_{\rm ref}$ \textit{(bottom)} in a disk column at the fiducial $r = 2 {\rm \ au}$, as compared to the final, relaxed state of $T_{\rm ref}$. As in Figure \ref{fig:test_self_shadow}, we plot the $\tau_r = 1$ optical surface in red and the $\beta = 1$ cooling surface in white. $T_{\rm ref}$, as well as $T_d$ and $T_g$, rapidly relax to a smooth equilibrium disk inside the $\beta = 1$ surface. Sound waves in the upper atmosphere---excited by the initial relaxation of the disk---are visible in $T_g$ due to the slow cooling rate at high altitude.}
    \label{fig:ssi_phenomenology}
\end{figure}

In Figure \ref{fig:test_self_shadow} we plot $T_d$ and $T_g$ from our realistic three-temperature simulation, as well as the matter temperature $T_{\rm ref}$ in the two-temperature reference simulation, at $t = \{0, 1, 10, 100\}$ y. In the reference simulation, the initial scale-height perturbations at the $\tau_r = 1$ surface dissipate rapidly, and are almost completely absent by $t = 10 {\rm \ y}$, reproducing the results of the equivalent two-temperature runs in \cite{MelonFuksman2022}. Our three-temperature realistic simulation, in which $\beta \approx 1$ at the $\tau_r = 1$ surface, shows a similar evolution, as $T_g$ still evolves too fast to achieve the smooth, columnar hydrostatic adaptation obtained in 1+1D and iterative approaches that reproduce the SSI. To better understand this temperature evolution, we plot in Figure \ref{fig:ssi_phenomenology} the time evolution of $T_{\rm d}$, $T_{\rm g}$, and $T_{\rm ref}$---normalized by the final, relaxed reference temperature $T_{\rm relax}$---in a fiducial disk column at $r = 2 {\rm \ au}$.

By $t = 10^2 {\rm y}$, the temperature profile within the $\tau_r = 1$ surface of the realistic simulation has largely reached the same smooth equilibrium as in the reference. The slow-cooling disk atmosphere in the realistic case---heated by waves emitted by the disk's initial relaxation and reflecting off the boundaries---is too low in density and optical depth to substantively alter this picture. It remains an open question to what extent three-temperature methods can sustain the SSI. In this regard, our numerical experiments with a dust-gas coupling time artificially increased by $100\times$ (described in Appendix \ref{sec:ssi_appendix}) show some promise, allowing SSI waves to survive and propagate inward for roughly $\sim10^2 {\rm y}$ before decaying.

\subsection{3D test: planet-driven spirals} \label{sec:spiral}
In recent years, high-resolution observations in NIR \citep[e.g.,][]{Benisty2015,Wagner2019,Xie2021} and in various molecular lines \citep[e.g.,][]{Teague2019,Woelfer2021} have revealed spiral structures in circumstellar disks. It is hypothesized that some fraction of these are driven by embedded, planetary-mass companions \citep{Goldreich1978,Goldreich1979,Goldreich1980,Goodman2001}.

Early simulations of these spirals \citep[e.g.,][]{Kley1999} often used two-dimensional, vertically averaged setups, with a locally isothermal equation of state. Subsequent advances in computational power made it feasible to run disk-planet interaction simulations in full 3D \citep[e.g.,][]{Fung2015,Dong2016}, relax the assumption of local isothermality \citep{Zhu2015}, and implement realistic vertical temperature structures \citep{Juhasz2018}---which, in turn, enabled meaningful comparisons of simulated spirals to observations in dust \citep{Dong2017} and gas \citep{Muley2021}. More recently, \cite{Bae21} used a parametrized cooling prescription to include the effects of matter-radiation interaction and gas-grain collisions. In what follows, we build on these previous works by modeling spirals with self-consistent, three-temperature radiation hydrodynamics.

For this test, we use the relaxed hydrostatic setup \texttt{dg3t0.1} from \cite{MelonFuksman2022} as a starting point. The dust properties, stellar parameters, and cooling prescription are the same as in section \ref{sec:ssi}, and we plot our initial condition, as well as the resulting contours for $\beta$ and $\tau_r$, in the lower panel of Figure \ref{fig:ssi_ic}. For this test, we use a $\hat{c} = 10^{-4} c$. Our simulation is run in 3D spherical coordinates, with radial boundaries at $r = \left[8, 100\right] \ {\rm au}$, polar boundaries at $\theta = \pi/2 + \left[-0.27, 0.27\right]$, and azimuthal boundaries at $\phi = \left[0, 2\pi\right]$; this simulation box is covered by $\{N_r, N_\theta, N_\phi\} = \{234, 54, 583\}$ cells, logarithmic in radius and uniform in azimuthal and polar angle. All fields are fixed to their initial values at the radial and azimuthal boundaries, and periodic at the azimuthal boundary.

In addition to the star and its gravitational potential $\Phi_*$, our simulation includes a $q \equiv M_p / M_* = 10^{-3}$ planet ($M_p \approx 1.05 M_J$), fixed on a circular Keplerian orbit with semimajor axis $a_p = 20 \ {\rm au}$; given our temperature profile, this is $q_{\rm th} = 5.5$ times the thermal mass. The planetary potential is given, following \cite{Klahr2006}, by
\begin{equation}
    \Phi_p = -\frac{GM_p}{r_p}\left\{\begin{matrix}
1 & {\rm if \ } r_p \geq r_s
\\ ([r_p/r_s]^4 - 2 [r_p/r_s]^3 + 2[r_p/r_s]) & {\rm if \ } r_p <r_s

\end{matrix}\right.
\end{equation}
where $r_p \equiv || \vec{x} - \vec{x}_p ||$ is the distance between the planet and a point in our domain, and $r_s$ is the smoothing length, implemented to avoid a singularity in the potential. In our simulations, we set $r_s = r_H/2$.

To avoid unphysically shocking the disk at early times, we introduce this companion at $t_i = 1 \ {\rm y}$ and grow it from zero to its final mass over the following $t_{\rm grow} = 10 \ {\rm y}$. We solve our equations in the planet's co-rotating frame, fixing its azimuthal coordinate as $\phi_p = \pi/4$. In this frame, the pattern speed of the spiral, and consequently of the non-axisymmetric radiation field, is zero; deviations would be of order $c_s \approx 10^{-6} c \approx 10^{-2}\hat{c}$. Numerical tests at $\hat{c}' = 2 \times 10^{-5} = \hat{c}/5$, not shown here, yield a well-converged spiral amplitude. To prevent the growth of large-scale disk instabilities such as the Rossby wave instability \cite[RWI][]{Lovelace2014}, we include a kinematic viscosity $\nu = \alpha c_s^2 \Omega^{-1}$, where $\alpha = 0.001$ is the \cite{Shakura1973} viscosity parameter. As in section \ref{sec:ssi}, we run a three-temperature realistic simulation, as well as an otherwise-identical reference simulation with $t_{\rm s} = (2\pi)^{-1} \times 10^{-10} {\rm \ y}$. 

\begin{figure}
    \centering
    \includegraphics[width=0.999\linewidth]{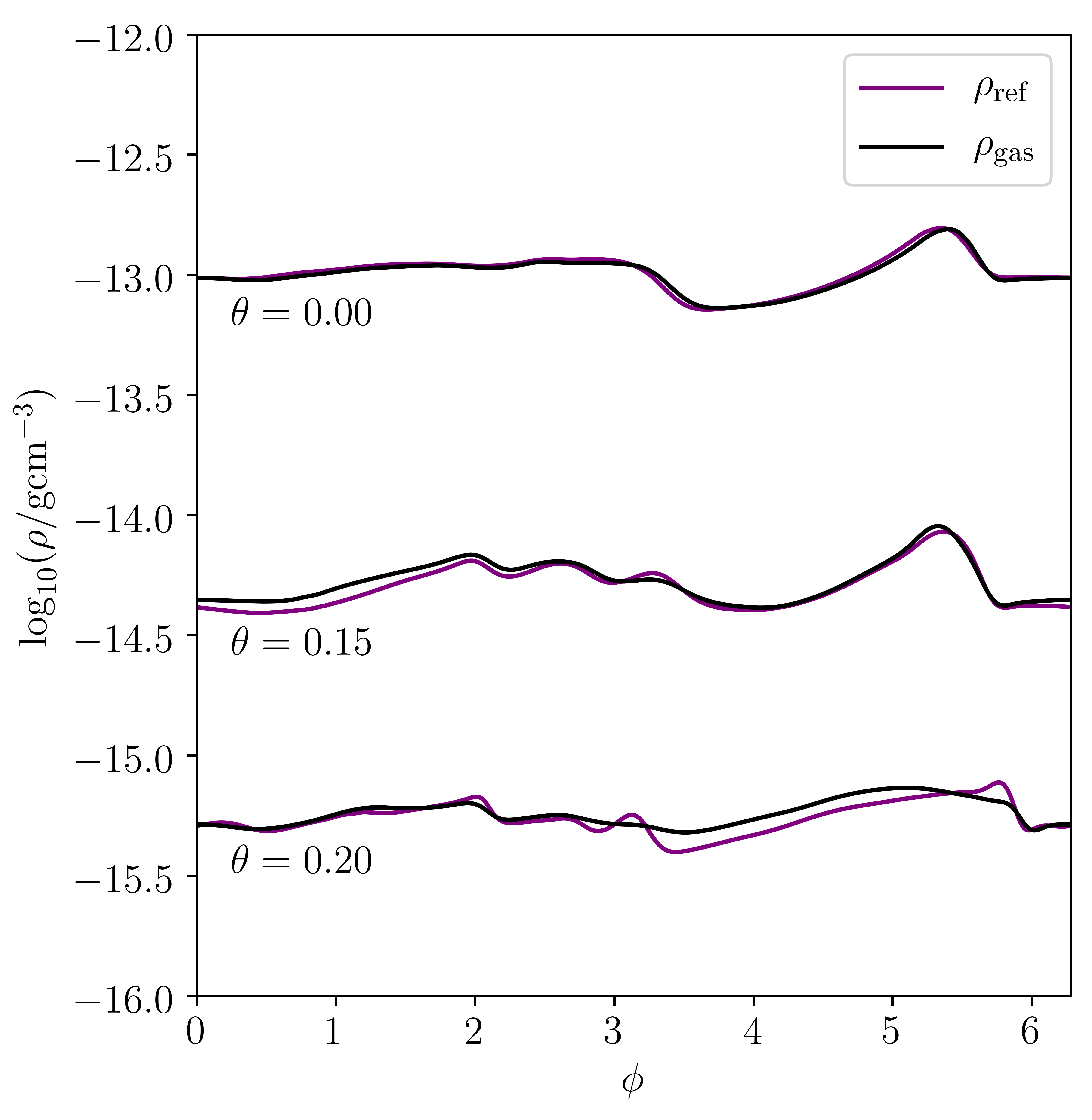}
    \caption{Density values for our three-temperature realistic ($\rho_{\rm gas}$) and reference ($\rho_{\rm ref}$) simulations, taken at the indicated $\theta$ values at a fixed $r = 30 {\rm \ au}$. Densities largely agree to within a few percent between both models, but the discrepancy rises to $\sim 20\%$ at the $\tau_r = 1$ surface, causing a mild impact on the temperature structure below. 
    }
    \label{fig:rho_spiral_cut}
\end{figure}

To keep focus on the spirals, we analyze the snapshot at $t_{\rm cut} = 500 \ {\rm y}$, before our super-thermal planet has had the chance to carve a gap in the disk. In Figure \ref{fig:rho_spiral_cut}, we present gas densities in our three-temperature realistic ($\rho_{\rm gas}$) and two-temperature reference ($\rho_{\rm ref}$) simulations, taken at $r = 30 {\rm \ au}$ and $\theta = \{0, 0.15, 0.2\}$. At $\theta = 0$, the primary and secondary Lindblad spirals are visible in both $\rho_{\rm gas}$ and $\rho_{\rm ref}$, and agree to within several percent. This is expected, because the three-temperature midplane $\beta \lesssim 10^{-2}$ and reference simulation global $\beta \lesssim 10^{-11}$ are both much shorter than the spiral-crossing timescale, $\beta_{\rm spiral} \approx (h/r)|1 - \Omega_p / \Omega_{30}|^{-1}$ \citep{Sturm2020} where $\Omega_{\rm 30}$ is the Keplerian frequency at 30 au. In the upper disk, the realistic $\beta \approx 1$ exceeds $\beta_{\rm spiral}$, while the reference $\beta$ remains unchanged. Due to this difference in thermal physics, the density structures become discrepant at the $\sim 20\%$ level, with Lindblad spirals being more sharply defined in $\rho_{\rm ref}$ than in $\rho_{\rm gas}$.

\begin{figure}
    \centering
        \includegraphics[width=0.959\linewidth]{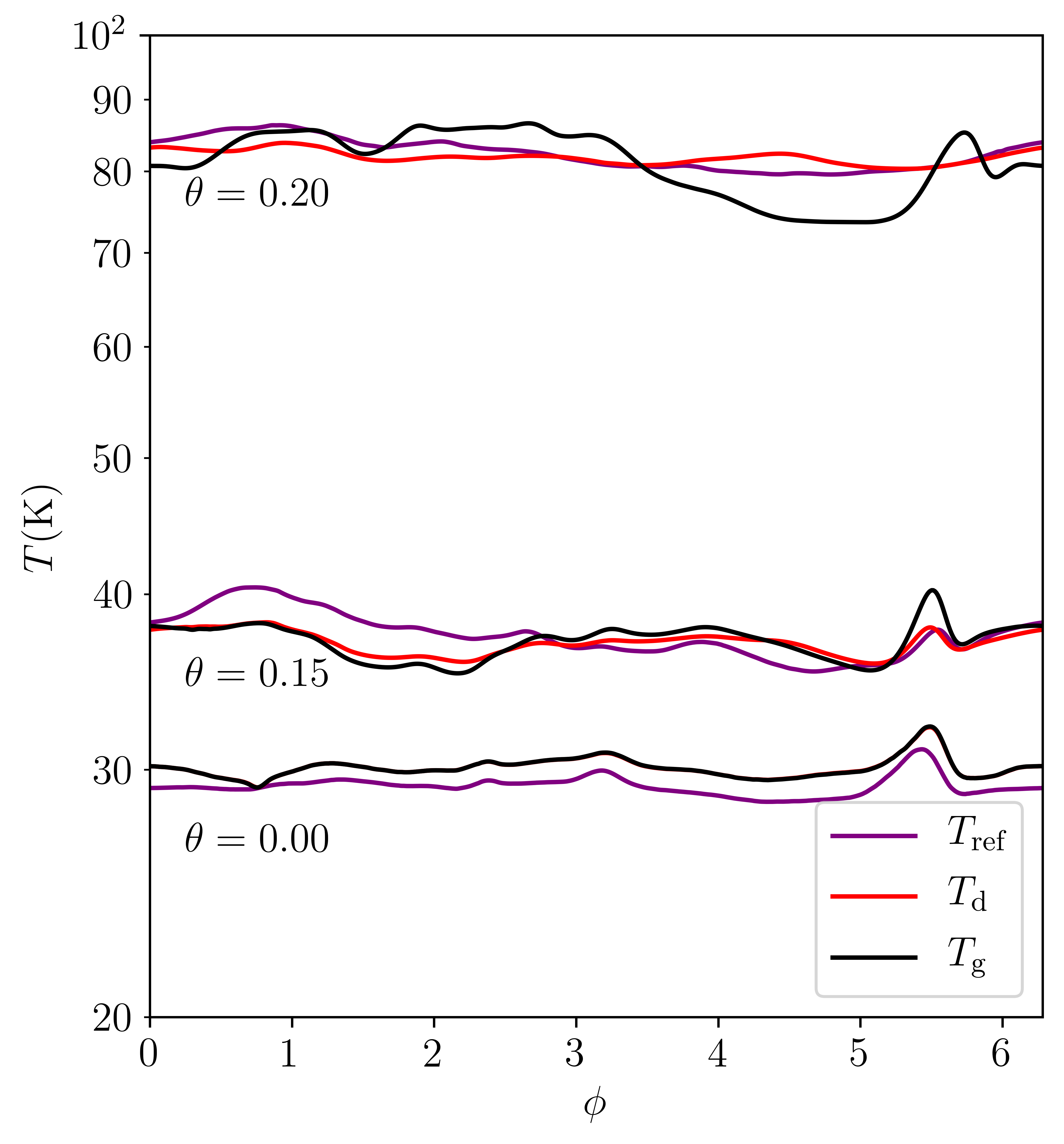}
    \caption{Temperatures in our three-temperature realistic ($T_{\rm gas}, T_{\rm dust}$) and reference ($T_{\rm ref}$) simulations, taken at the same cuts as in Figure \ref{fig:rho_spiral_cut}. In the midplane, where coupling times are short, $T_{\rm gas} = T_{\rm dust}$, but in the disk atmosphere, where coupling is weaker, there is a noticeable ($\sim 10\%$) discrepancy between the two: $T_{\rm dust}$ reflects the stellar irradiation field, while $T_{\rm gas}$ is driven by the $P dV$ work done by the planetary spiral.
    }
    \label{fig:temp_spiral_cut}
\end{figure}

In Figure \ref{fig:temp_spiral_cut}, we plot temperatures at the same cuts studied in Figure \ref{fig:rho_spiral_cut}. In the midplane of our three-temperature simulation, strong dust-gas coupling ensures that $T_d$ and $T_g$ are equal. At $\theta = 0.15$, dust-gas coupling is somewhat weaker, with a $\beta \approx 10^{-1} \approx \beta_{\rm spiral}$, and differences between $T_d$ and $T_g$ are visible throughout the azimuthal extent of our cut. High in the disk atmosphere, at $\theta = 0.2$, the normalized cooling time is $\beta \approx 1$, much longer than the spiral-crossing timescale for gas. Here, we observe a spiral feature in $T_g$ with an amplitude ${\sim}10\%$---comparable to that found the $\beta = 1$, $M_p = 400 M_\oplus$ spiral simulation in \cite{Muley2021}---which is not visible in $T_d$ or $T_{\rm ref}$.

At $\theta = 0$, we observe a slight temperature dip centered at $\phi = \phi_p = \pi/4$, while at $\theta = 0.15$ and $\theta = 0.2$, we find a temperature bump at the same azimuthal position. This is ultimately caused by the gravitational influence of the planet, which lowers the $\tau_r = 1$ surface overhead and thus changes the illumination of material in the radial band $\phi \approx \phi_p$, $r \gtrsim r_p$. Such a feature was also observed in the Monte Carlo radiative transfer post-processing of \cite{Muley2021}'s hydrodynamical simulations.

\begin{figure*}
    \centering
    \includegraphics[width=0.95\textwidth]{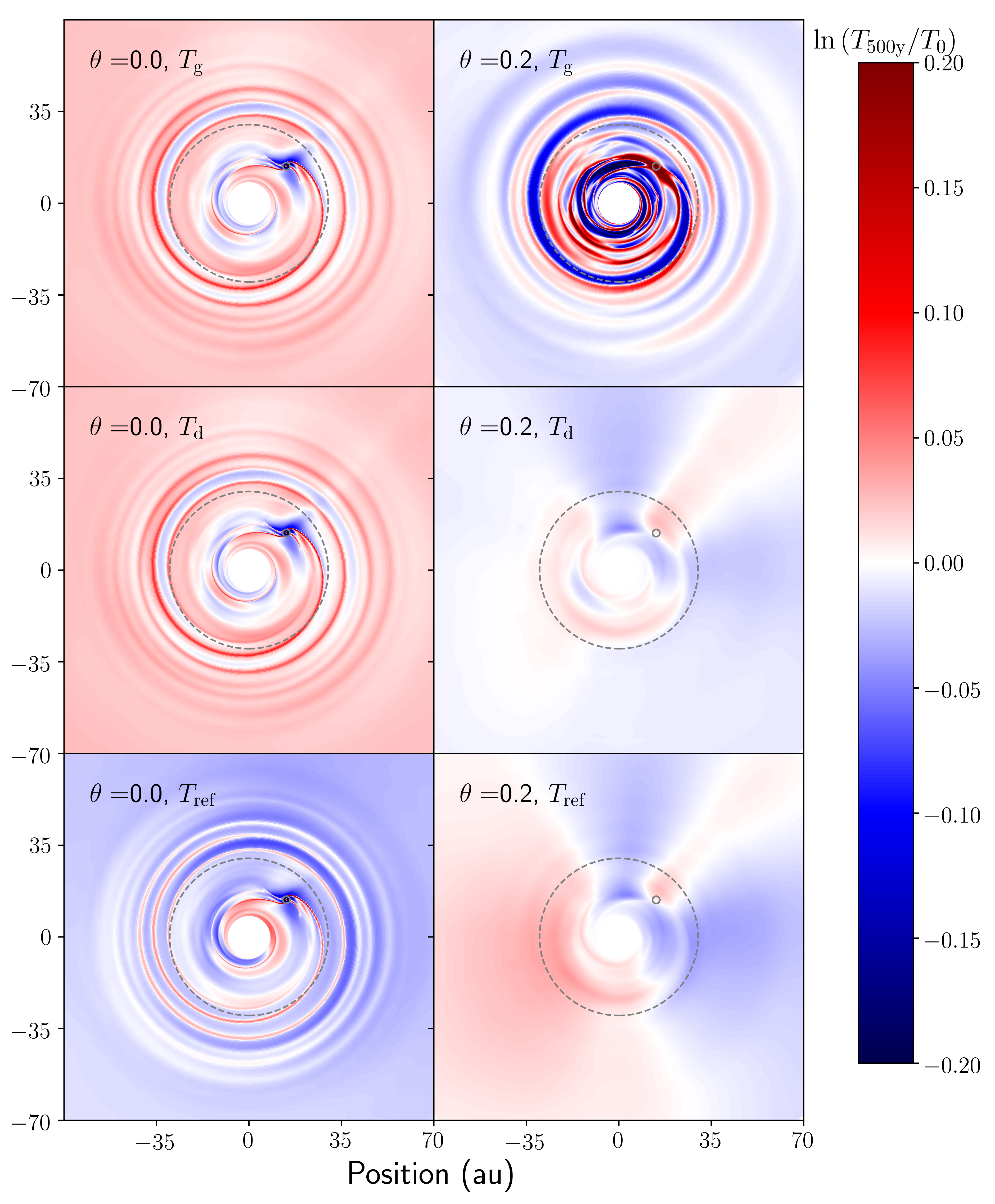}
    \caption{Relative difference between the temperature at $t_{\rm cut} = 500$ y and the initial profile, in polar cuts at $\theta = 0.0$ and $\theta = 0.2$. The morphology of $T_d$ and $T_g$ agree in the midplane, but diverge at higher altitude, while the morphology of $T_d$ and $T_{\rm ref}$ agree everywhere. In each panel, we indicate the Hill sphere of the planet with a small \textit{solid grey} circle, and our cuts in figures \ref{fig:rho_spiral_cut} and \ref{fig:temp_spiral_cut} with a large \textit{dotted grey} circle.}
    \label{fig:temp_diff_spiral_cut}
\end{figure*}

To provide a more visual understanding of these results, we present 2D polar cuts of $T_d$, $T_g$, and $T_{\rm ref}$, at $\theta = 0.0$ and $\theta = 0.2$, in Figure \ref{fig:temp_diff_spiral_cut}. We find, as in our 1D quantitative plots, that the morphology of $T_{\rm ref}$ largely agrees with that of $T_{d}$, despite substantial differences in $\beta$ between and within each simulation, while $T_g$ and $T_d$ only agree with one another when $\beta < \beta_{\rm spiral}$. We interpret this to imply that $T_d$ is set by the very short timescales for \textit{dust} collisional energy exchange $t_{\rm c, dust} = r_{\rm gd}^{-1} t_c$ (following equation \ref{eq:tcool_d}) and for radiative cooling, either vertically or in-plane \citep{Miranda2020,Ziampras2020}. How $T_g$, in turn, reacts to the dust depends on the much longer local gas cooling time $t_c = \beta/\Omega$---a result that could affect the observational signatures disk-planet interaction produces in gas and dust. We note, in addition, a slight discrepancy between the background $T_{\rm dust}$ and $T_{\rm ref}$, with the latter being systematically lower throughout the disk; this is because spirals are more vertically extended in the two-temperature limit, and so cast stronger shadows throughout the disk. We intend to investigate these issues further in a future study.

\section{Conclusion}\label{sec:conclusion}
We have developed an implicit-explicit (IMEX) numerical scheme for the PLUTO hydrodynamics code to self-consistently model energy exchange between gas, dust, and radiation. The implicit substep, consisting of nonrelativistic matter-radiation interaction source terms, is solved iteratively with a multidimensional Newton-Raphson method, whereas the explicit substep, consisting of radiation transport and relativistic interaction source terms, is solved using a Godunov scheme, following \cite{MelonFuksman2019}. Because this explicit part is Courant-limited, we use the reduced speed of light approximation, as implemented by \cite{MelonFuksman2021}, to improve computational efficiency. This gives physically reasonable results, so long as the reduced speed of light $\hat{c}$ is much larger than all other relevant velocities in the problem.

In our 0D matter-radiation coupling test, we demonstrate the ability of our scheme to reproduce the ``two-temperature'' regime of traditional radiation hydrodynamics, where dust and gas are well-coupled. For the 1D \cite{Ensman1994} shock tube test, we relax this assumption, and investigate the behavior of gas, dust, and radiation temperatures in different coupling regimes. Our subsequent tests illustrate the applicability of three-temperature radiation hydrodynamics to active problems in circumstellar disk research, including realistic opacities and dust-gas coupling times computed from local temperature and density. Our 2D test shows that the self-shadowing instability---which arises in 1+1D time-dependent models with parametrized disk scale height, as well as 2D iterative techniques that alternate the calculation of radiative and hydrostatic equilibrium---decays rapidly in the two-temperature limit, as well as with realistic three-temperature coupling times, in line with previous expectations \citep{MelonFuksman2022}. However, an increased cooling time (Appendix \ref{sec:ssi_appendix}) allows dust temperatures in most of the disk column to rapidly adapt to stellar irradiation before the gas can react---which better satisfies the key assumptions of the aforementioned iterative and 1D-hydrostatic methods, and so allows SSI-type scale-height perturbations to survive for somewhat longer before decaying. In our 3D spiral simulations, dust temperatures are remarkably consistent between the two- and three-temperature regimes, but gas temperatures disagree in the upper atmosphere---a result with implications for observations of spirals in gas tracers such as $^{12}$CO. Future potential applications of our three-temperature scheme include studies of circumplanetary disk accretion shocks, planetary gap opening, photoevaporation, and hydrodynamic/MHD instabilities such as the vertical shear instability, among many others.

Our scheme makes the simplifications that dust and gas move at the same velocity, and that dust inertia is negligible (${\rm St} \equiv t_s \Omega \ll 1$, $f_{\rm d} \ll 1$). While these are reasonable assumptions for the sub-$\mu$m grains that supply most of a disk's total opacity and set its overall thermal structure, they are invalid for the millimeter grains that concentrate at pressure maxima and produce much of its continuum emission. Including such grains would require accounting for multispecies momentum transfer \revbf{including back-reaction from the gas} \citep[using schemes such as those detailed in e.g.,][]{BenitezLlambay2019,Huang2022} and thermal coupling
. While such a method would be immensely helpful in interpreting continuum observations of gaps, rings, and spirals, it would be computationally demanding and require extensive testing beyond the scope of this work.

\begin{acknowledgements}
We thank Remo Burn, Cornelis Dullemond, Mario Flock, Thomas Henning, Carolin Kimmig, Giancarlo Mattia, Thomas Rometsch, and Prakruti Sudarshan for useful discussions. \revbf{We are also grateful to the anonymous referee, whose report helped improve the quality of this paper.} All numerical simulations were performed on the Cobra cluster of the Max Planck Society and the Vera Cluster of the Max Planck Institut f\"ur Astronomie, both hosted by the Max Planck Computing and Data Facility (MPCDF) in Garching bei München. The research of J.D.M.F. and H.K. is supported by the German Science Foundation (DFG) under the priority program SPP 1992: ``Exoplanet Diversity" under contract KL 1469/16-1/2. 
\end{acknowledgements}


\begin{thebibliography}{81}
\expandafter\ifx\csname natexlab\endcsname\relax\def\natexlab#1{#1}\fi

\bibitem[{{Bae} {et~al.}(2021){Bae}, {Teague}, \& {Zhu}}]{Bae21}
{Bae}, J., {Teague}, R., \& {Zhu}, Z. 2021, \apj, 912, 56

\bibitem[{{Bate} \& {Keto}(2015)}]{Bate2015}
{Bate}, M.~R. \& {Keto}, E.~R. 2015, \mnras, 449, 2643

\bibitem[{{Benisty} {et~al.}(2015){Benisty}, {Juhasz}, {Boccaletti},
  {Avenhaus}, {Milli}, {Thalmann}, {Dominik}, {Pinilla}, {Buenzli}, {Pohl},
  {Beuzit}, {Birnstiel}, {de Boer}, {Bonnefoy}, {Chauvin}, {Christiaens},
  {Garufi}, {Grady}, {Henning}, {Huelamo}, {Isella}, {Langlois}, {M{\'e}nard},
  {Mouillet}, {Olofsson}, {Pantin}, {Pinte}, \& {Pueyo}}]{Benisty2015}
{Benisty}, M., {Juhasz}, A., {Boccaletti}, A., {et~al.} 2015, \aap, 578, L6

\bibitem[{{Ben{\'\i}tez-Llambay} {et~al.}(2019){Ben{\'\i}tez-Llambay}, {Krapp},
  \& {Pessah}}]{BenitezLlambay2019}
{Ben{\'\i}tez-Llambay}, P., {Krapp}, L., \& {Pessah}, M.~E. 2019, \apjs, 241,
  25

\bibitem[{{Binkert} {et~al.}(2023){Binkert}, {Szul{\'a}gyi}, \&
  {Birnstiel}}]{Binkert2023}
{Binkert}, F., {Szul{\'a}gyi}, J., \& {Birnstiel}, T. 2023, \mnras

\bibitem[{{Bitsch} {et~al.}(2013){Bitsch}, {Crida}, {Morbidelli}, {Kley}, \&
  {Dobbs-Dixon}}]{Bitsch2013a}
{Bitsch}, B., {Crida}, A., {Morbidelli}, A., {Kley}, W., \& {Dobbs-Dixon}, I.
  2013, \aap, 549, A124

\bibitem[{{Bruderer}(2013)}]{Bruderer2013}
{Bruderer}, S. 2013, \aap, 559, A46

\bibitem[{{Bruderer} {et~al.}(2012){Bruderer}, {van Dishoeck}, {Doty}, \&
  {Herczeg}}]{Bruderer2012}
{Bruderer}, S., {van Dishoeck}, E.~F., {Doty}, S.~D., \& {Herczeg}, G.~J. 2012,
  \aap, 541, A91

\bibitem[{{Burke} \& {Hollenbach}(1983)}]{BurkeHollenbach83}
{Burke}, J.~R. \& {Hollenbach}, D.~J. 1983, \apj, 265, 223

\bibitem[{{Calahan} {et~al.}(2021){Calahan}, {Bergin}, {Zhang}, {Schwarz},
  {{\"O}berg}, {Guzm{\'a}n}, {Walsh}, {Aikawa}, {Alarc{\'o}n}, {Andrews},
  {Bae}, {Bergner}, {Booth}, {Bosman}, {Cataldi}, {Czekala}, {Huang}, {Ilee},
  {Law}, {Le Gal}, {Long}, {Loomis}, {M{\'e}nard}, {Nomura}, {Qi}, {Teague},
  {van't Hoff}, {Wilner}, \& {Yamato}}]{Calahan2021MAPS}
{Calahan}, J.~K., {Bergin}, E.~A., {Zhang}, K., {et~al.} 2021, \apjs, 257, 17

\bibitem[{{Chiang} \& {Goldreich}(1997)}]{Chiang1997}
{Chiang}, E.~I. \& {Goldreich}, P. 1997, \apj, 490, 368

\bibitem[{{D'Alessio} {et~al.}(1999){D'Alessio}, {Cant{\'o}}, {Hartmann},
  {Calvet}, \& {Lizano}}]{DAlessio1999}
{D'Alessio}, P., {Cant{\'o}}, J., {Hartmann}, L., {Calvet}, N., \& {Lizano}, S.
  1999, \apj, 511, 896

\bibitem[{{Dong} \& {Fung}(2017)}]{Dong2017}
{Dong}, R. \& {Fung}, J. 2017, \apj, 835, 38

\bibitem[{{Dong} {et~al.}(2016){Dong}, {Fung}, \& {Chiang}}]{Dong2016}
{Dong}, R., {Fung}, J., \& {Chiang}, E. 2016, \apj, 826, 75

\bibitem[{{Dullemond}(2000)}]{Dullemond2000}
{Dullemond}, C.~P. 2000, \aap, 361, L17

\bibitem[{{Dullemond} {et~al.}(2007){Dullemond}, {Hollenbach}, {Kamp}, \&
  {D'Alessio}}]{Dullemond2007}
{Dullemond}, C.~P., {Hollenbach}, D., {Kamp}, I., \& {D'Alessio}, P. 2007, in
  Protostars and Planets V, ed. B.~{Reipurth}, D.~{Jewitt}, \& K.~{Keil}, 555

\bibitem[{{Dullemond} {et~al.}(2012){Dullemond}, {Juhasz}, {Pohl}, {Sereshti},
  {Shetty}, {Peters}, {Commercon}, \& {Flock}}]{Dullemond2012}
{Dullemond}, C.~P., {Juhasz}, A., {Pohl}, A., {et~al.} 2012, {RADMC-3D: A
  multi-purpose radiative transfer tool}, Astrophysics Source Code Library,
  record ascl:1202.015

\bibitem[{{Ensman}(1994)}]{Ensman1994}
{Ensman}, L. 1994, \apj, 424, 275

\bibitem[{{Flock} {et~al.}(2013){Flock}, {Fromang}, {Gonz{\'a}lez}, \&
  {Commer{\c{c}}on}}]{Flock2013}
{Flock}, M., {Fromang}, S., {Gonz{\'a}lez}, M., \& {Commer{\c{c}}on}, B. 2013,
  \aap, 560, A43

\bibitem[{{Flock} {et~al.}(2020){Flock}, {Turner}, {Nelson}, {Lyra}, {Manger},
  \& {Klahr}}]{Flock2020}
{Flock}, M., {Turner}, N.~J., {Nelson}, R.~P., {et~al.} 2020, \apj, 897, 155

\bibitem[{{Fung} \& {Dong}(2015)}]{Fung2015}
{Fung}, J. \& {Dong}, R. 2015, \apjl, 815, L21

\bibitem[{{Fung} {et~al.}(2019){Fung}, {Zhu}, \& {Chiang}}]{Fung2019}
{Fung}, J., {Zhu}, Z., \& {Chiang}, E. 2019, \apj, 887, 152

\bibitem[{{Gnedin} \& {Abel}(2001)}]{Gnedin2001}
{Gnedin}, N.~Y. \& {Abel}, T. 2001, \na, 6, 437

\bibitem[{{Goldreich} \& {Tremaine}(1978)}]{Goldreich1978}
{Goldreich}, P. \& {Tremaine}, S. 1978, \apj, 222, 850

\bibitem[{{Goldreich} \& {Tremaine}(1979)}]{Goldreich1979}
{Goldreich}, P. \& {Tremaine}, S. 1979, \apj, 233, 857

\bibitem[{{Goldreich} \& {Tremaine}(1980)}]{Goldreich1980}
{Goldreich}, P. \& {Tremaine}, S. 1980, \apj, 241, 425

\bibitem[{{Goodman} \& {Rafikov}(2001)}]{Goodman2001}
{Goodman}, J. \& {Rafikov}, R.~R. 2001, \apj, 552, 793

\bibitem[{{Hu} {et~al.}(2023){Hu}, {Li}, {Wang}, {Zhu}, \& {Bae}}]{Hu2023}
{Hu}, X., {Li}, Z.-Y., {Wang}, L., {Zhu}, Z., \& {Bae}, J. 2023, arXiv
  e-prints, arXiv:2304.05972

\bibitem[{{Huang} \& {Bai}(2022)}]{Huang2022}
{Huang}, P. \& {Bai}, X.-N. 2022, \apjs, 262, 11

\bibitem[{{Huang} \& {Yu}(2022)}]{HuangYu2022}
{Huang}, S. \& {Yu}, C. 2022, \mnras, 514, 1733

\bibitem[{{Hubeny}(1990)}]{Hubeny1990}
{Hubeny}, I. 1990, \apj, 351, 632

\bibitem[{{Juh{\'a}sz} \& {Rosotti}(2018)}]{Juhasz2018}
{Juh{\'a}sz}, A. \& {Rosotti}, G.~P. 2018, \mnras, 474, L32

\bibitem[{{Keppler} {et~al.}(2018){Keppler}, {Benisty}, {M{\"u}ller},
  {Henning}, {van Boekel}, {Cantalloube}, {Ginski}, {van Holstein}, {Maire},
  {Pohl}, {Samland}, {Avenhaus}, {Baudino}, {Boccaletti}, {de Boer},
  {Bonnefoy}, {Chauvin}, {Desidera}, {Langlois}, {Lazzoni}, {Marleau},
  {Mordasini}, {Pawellek}, {Stolker}, {Vigan}, {Zurlo}, {Birnstiel},
  {Brandner}, {Feldt}, {Flock}, {Girard}, {Gratton}, {Hagelberg}, {Isella},
  {Janson}, {Juhasz}, {Kemmer}, {Kral}, {Lagrange}, {Launhardt}, {Matter},
  {M{\'e}nard}, {Milli}, {Molli{\`e}re}, {Olofsson}, {P{\'e}rez}, {Pinilla},
  {Pinte}, {Quanz}, {Schmidt}, {Udry}, {Wahhaj}, {Williams}, {Buenzli},
  {Cudel}, {Dominik}, {Galicher}, {Kasper}, {Lannier}, {Mesa}, {Mouillet},
  {Peretti}, {Perrot}, {Salter}, {Sissa}, {Wildi}, {Abe}, {Antichi},
  {Augereau}, {Baruffolo}, {Baudoz}, {Bazzon}, {Beuzit}, {Blanchard}, {Brems},
  {Buey}, {De Caprio}, {Carbillet}, {Carle}, {Cascone}, {Cheetham}, {Claudi},
  {Costille}, {Delboulb{\'e}}, {Dohlen}, {Fantinel}, {Feautrier}, {Fusco},
  {Giro}, {Gluck}, {Gry}, {Hubin}, {Hugot}, {Jaquet}, {Le Mignant}, {Llored},
  {Madec}, {Magnard}, {Martinez}, {Maurel}, {Meyer}, {M{\"o}ller-Nilsson},
  {Moulin}, {Mugnier}, {Orign{\'e}}, {Pavlov}, {Perret}, {Petit}, {Pragt},
  {Puget}, {Rabou}, {Ramos}, {Rigal}, {Rochat}, {Roelfsema}, {Rousset}, {Roux},
  {Salasnich}, {Sauvage}, {Sevin}, {Soenke}, {Stadler}, {Suarez}, {Turatto}, \&
  {Weber}}]{Keppler2018}
{Keppler}, M., {Benisty}, M., {M{\"u}ller}, A., {et~al.} 2018, \aap, 617, A44

\bibitem[{{Klahr} \& {Kley}(2006)}]{Klahr2006}
{Klahr}, H. \& {Kley}, W. 2006, \aap, 445, 747

\bibitem[{{Kley}(1999)}]{Kley1999}
{Kley}, W. 1999, \mnras, 303, 696

\bibitem[{{Kley} {et~al.}(2009){Kley}, {Bitsch}, \& {Klahr}}]{Kley2009}
{Kley}, W., {Bitsch}, B., \& {Klahr}, H. 2009, \aap, 506, 971

\bibitem[{{Krieger} \& {Wolf}(2020)}]{Krieger2020}
{Krieger}, A. \& {Wolf}, S. 2020, \aap, 635, A148

\bibitem[{{Krieger} \& {Wolf}(2022)}]{Krieger2022}
{Krieger}, A. \& {Wolf}, S. 2022, \aap, 662, A99

\bibitem[{{Levermore}(1984)}]{Levermore1984}
{Levermore}, C.~D. 1984, \jqsrt, 31, 149

\bibitem[{{Levermore} \& {Pomraning}(1981)}]{Levermore1981}
{Levermore}, C.~D. \& {Pomraning}, G.~C. 1981, \apj, 248, 321

\bibitem[{{Lovelace} \& {Romanova}(2014)}]{Lovelace2014}
{Lovelace}, R.~V.~E. \& {Romanova}, M.~M. 2014, Fluid Dynamics Research, 46,
  041401

\bibitem[{{Malygin} {et~al.}(2017){Malygin}, {Klahr}, {Semenov}, {Henning}, \&
  {Dullemond}}]{Malygin2017}
{Malygin}, M.~G., {Klahr}, H., {Semenov}, D., {Henning}, T., \& {Dullemond},
  C.~P. 2017, \aap, 605, A30

\bibitem[{{Manger} {et~al.}(2021){Manger}, {Pfeil}, \& {Klahr}}]{Manger2021}
{Manger}, N., {Pfeil}, T., \& {Klahr}, H. 2021, \mnras, 508, 5402

\bibitem[{{Mathis} {et~al.}(1977){Mathis}, {Rumpl}, \& {Nordsieck}}]{MRN77}
{Mathis}, J.~S., {Rumpl}, W., \& {Nordsieck}, K.~H. 1977, \apj, 217, 425

\bibitem[{{Melon Fuksman} {et~al.}(2023){Melon Fuksman}, {Flock}, \&
  {Klahr}}]{MelonFuksman2023}
{Melon Fuksman}, J.~D., {Flock}, M., \& {Klahr}, H. 2023, \aap, submitted

\bibitem[{{Melon Fuksman} \& {Klahr}(2022)}]{MelonFuksman2022}
{Melon Fuksman}, J.~D. \& {Klahr}, H. 2022, \apj, 936, 16

\bibitem[{{Melon Fuksman} {et~al.}(2021){Melon Fuksman}, {Klahr}, {Flock}, \&
  {Mignone}}]{MelonFuksman2021}
{Melon Fuksman}, J.~D., {Klahr}, H., {Flock}, M., \& {Mignone}, A. 2021, \apj,
  906, 78

\bibitem[{{Melon Fuksman} \& {Mignone}(2019)}]{MelonFuksman2019}
{Melon Fuksman}, J.~D. \& {Mignone}, A. 2019, \apjs, 242, 20

\bibitem[{{Mignone} {et~al.}(2007){Mignone}, {Bodo}, {Massaglia}, {Matsakos},
  {Tesileanu}, {Zanni}, \& {Ferrari}}]{Mignone2007}
{Mignone}, A., {Bodo}, G., {Massaglia}, S., {et~al.} 2007, \apjs, 170, 228

\bibitem[{Miranda \& Rafikov(2019)}]{Miranda2019}
Miranda, R. \& Rafikov, R.~R. 2019, The Astrophysical Journal Letters, 878, L9

\bibitem[{{Miranda} \& {Rafikov}(2020{\natexlab{a}})}]{Miranda2020}
{Miranda}, R. \& {Rafikov}, R.~R. 2020{\natexlab{a}}, \apj, 904, 121

\bibitem[{{Miranda} \& {Rafikov}(2020{\natexlab{b}})}]{Miranda2020b}
{Miranda}, R. \& {Rafikov}, R.~R. 2020{\natexlab{b}}, \apj, 892, 65

\bibitem[{{Muley} {et~al.}(2021){Muley}, {Dong}, \& {Fung}}]{Muley2021}
{Muley}, D., {Dong}, R., \& {Fung}, J. 2021, \aj, 162, 129

\bibitem[{Pareschi \& Russo(2005)}]{Pareschi2005}
Pareschi, L. \& Russo, G. 2005, Journal of Scientific Computing, 25, 129

\bibitem[{{Pavlyuchenkov} \& {Zhilkin}(2013)}]{Pavlyuchenkov2013}
{Pavlyuchenkov}, Y.~N. \& {Zhilkin}, A.~G. 2013, Astronomy Reports, 57, 641

\bibitem[{{Pavlyuchenkov} {et~al.}(2015){Pavlyuchenkov}, {Zhilkin}, {Vorobyov},
  \& {Fateeva}}]{Pavlyuchenkov2015}
{Pavlyuchenkov}, Y.~N., {Zhilkin}, A.~G., {Vorobyov}, E.~I., \& {Fateeva},
  A.~M. 2015, Astronomy Reports, 59, 133

\bibitem[{{Pierens} \& {Lin}(2018)}]{Pierens2018}
{Pierens}, A. \& {Lin}, M.-K. 2018, \mnras, 479, 4878

\bibitem[{{Pinte} {et~al.}(2022){Pinte}, {Teague}, {Flaherty}, {Hall},
  {Facchini}, \& {Casassus}}]{Pinte2022}
{Pinte}, C., {Teague}, R., {Flaherty}, K., {et~al.} 2022, arXiv e-prints,
  arXiv:2203.09528

\bibitem[{{Ren} {et~al.}(2018){Ren}, {Dong}, {Esposito}, {Pueyo}, {Debes},
  {Poteet}, {Choquet}, {Benisty}, {Chiang}, {Grady}, {Hines}, {Schneider}, \&
  {Soummer}}]{Dong2018}
{Ren}, B., {Dong}, R., {Esposito}, T.~M., {et~al.} 2018, \apjl, 857, L9

\bibitem[{{Shakura} \& {Sunyaev}(1973)}]{Shakura1973}
{Shakura}, N.~I. \& {Sunyaev}, R.~A. 1973, \aap, 24, 337

\bibitem[{{Skinner} \& {Ostriker}(2013)}]{Skinner2013}
{Skinner}, M.~A. \& {Ostriker}, E.~C. 2013, \apjs, 206, 21

\bibitem[{{Speedie} {et~al.}(2022){Speedie}, {Booth}, \& {Dong}}]{Speedie2022}
{Speedie}, J., {Booth}, R.~A., \& {Dong}, R. 2022, \apj, 930, 40

\bibitem[{{Sturm} {et~al.}(2020){Sturm}, {Rosotti}, \& {Dominik}}]{Sturm2020}
{Sturm}, J.~A., {Rosotti}, G.~P., \& {Dominik}, C. 2020, \aap, 643, A92

\bibitem[{{Tarczay-Neh{\'e}z} {et~al.}(2020){Tarczay-Neh{\'e}z}, {Reg{\'a}ly},
  \& {Vorobyov}}]{TarczayNehez2020}
{Tarczay-Neh{\'e}z}, D., {Reg{\'a}ly}, Z., \& {Vorobyov}, E. 2020, \mnras, 493,
  3014

\bibitem[{{Teague} {et~al.}(2019){Teague}, {Bae}, {Huang}, \&
  {Bergin}}]{Teague2019}
{Teague}, R., {Bae}, J., {Huang}, J., \& {Bergin}, E.~A. 2019, \apjl, 884, L56

\bibitem[{{Ueda} {et~al.}(2021){Ueda}, {Flock}, \& {Birnstiel}}]{Ueda2021}
{Ueda}, T., {Flock}, M., \& {Birnstiel}, T. 2021, \apjl, 914, L38

\bibitem[{{Ueda} {et~al.}(2019){Ueda}, {Flock}, \& {Okuzumi}}]{Ueda2019}
{Ueda}, T., {Flock}, M., \& {Okuzumi}, S. 2019, \apj, 871, 10

\bibitem[{{Vorobyov} {et~al.}(2020){Vorobyov}, {Matsukoba}, {Omukai}, \&
  {Guedel}}]{Vorobyov2020}
{Vorobyov}, E.~I., {Matsukoba}, R., {Omukai}, K., \& {Guedel}, M. 2020, \aap,
  638, A102

\bibitem[{{Wagner} {et~al.}(2019){Wagner}, {Stone}, {Spalding}, {Apai}, {Dong},
  {Ertel}, {Leisenring}, \& {Webster}}]{Wagner2019}
{Wagner}, K., {Stone}, J.~M., {Spalding}, E., {et~al.} 2019, \apj, 882, 20

\bibitem[{{Wang} {et~al.}(2019){Wang}, {Bai}, \& {Goodman}}]{Wang2019}
{Wang}, L., {Bai}, X.-N., \& {Goodman}, J. 2019, \apj, 874, 90

\bibitem[{{Wang} \& {Goodman}(2017)}]{Wang2017}
{Wang}, L. \& {Goodman}, J. 2017, \apj, 847, 11

\bibitem[{{Watanabe} \& {Lin}(2008)}]{Watanabe2008}
{Watanabe}, S.-i. \& {Lin}, D.~N.~C. 2008, \apj, 672, 1183

\bibitem[{{Whitney} {et~al.}(2013){Whitney}, {Robitaille}, {Bjorkman}, {Dong},
  {Wolff}, {Wood}, \& {Honor}}]{Whitney2013}
{Whitney}, B.~A., {Robitaille}, T.~P., {Bjorkman}, J.~E., {et~al.} 2013, \apjs,
  207, 30

\bibitem[{{Woitke} {et~al.}(2009){Woitke}, {Kamp}, \& {Thi}}]{Woitke2009}
{Woitke}, P., {Kamp}, I., \& {Thi}, W.~F. 2009, \aap, 501, 383

\bibitem[{{W{\"o}lfer} {et~al.}(2021){W{\"o}lfer}, {Facchini}, {Kurtovic},
  {Teague}, {van Dishoeck}, {Benisty}, {Ercolano}, {Lodato}, {Miotello},
  {Rosotti}, {Testi}, \& {Ubeira Gabellini}}]{Woelfer2021}
{W{\"o}lfer}, L., {Facchini}, S., {Kurtovic}, N.~T., {et~al.} 2021, \aap, 648,
  A19

\bibitem[{{Wu} \& {Lithwick}(2021)}]{Wu2021}
{Wu}, Y. \& {Lithwick}, Y. 2021, \apj, 923, 123

\bibitem[{{Xie} {et~al.}(2021){Xie}, {Ren}, {Dong}, {Pueyo}, {Ruffio}, {Fang},
  {Mawet}, \& {Stolker}}]{Xie2021}
{Xie}, C., {Ren}, B., {Dong}, R., {et~al.} 2021, \apjl, 906, L9

\bibitem[{{Yamaleev} \& {Carpenter}(2009)}]{Yamaleev2009}
{Yamaleev}, N.~K. \& {Carpenter}, M.~H. 2009, Journal of Computational Physics,
  228, 3025

\bibitem[{{Zhang} \& {Zhu}(2020)}]{Zhang2020}
{Zhang}, S. \& {Zhu}, Z. 2020, \mnras, 493, 2287

\bibitem[{{Zhu} {et~al.}(2015){Zhu}, {Dong}, {Stone}, \& {Rafikov}}]{Zhu2015}
{Zhu}, Z., {Dong}, R., {Stone}, J.~M., \& {Rafikov}, R.~R. 2015, \apj, 813, 88

\bibitem[{{Ziampras} {et~al.}(2020){Ziampras}, {Kley}, \&
  {Dullemond}}]{Ziampras2020}
{Ziampras}, A., {Kley}, W., \& {Dullemond}, C.~P. 2020, \aap, 637, A50

\end{thebibliography}


\begin{appendix}

\section{Newton-Raphson iterations}\label{sec:newtonmethod}
We rearrange \cref{eq:edeq,eq:egeq} as a vector of quantities $\vec{P}(\vec{\xi}_{
\rm sub}^{i+1}) \equiv (P_d, P_g)^\top$, where $\vec{\xi}_{
\rm sub}^{i+1} \equiv (\xi_d^{i+1}, \xi_g^{i+1})$. 
\begin{equation}
\begin{split}
    P_d &= \xi_d^{i} - \xi_d^{i+1} - t_c^{-1} \left[r_{gd} \xi_d^{i+1} - \xi_g^{i+1}\right] \delta t \\
    & - c\rho\kappa_d^{i+1}f_d\left[a_r (T_d^{i+1})^4 - \xi_r^{i+1}\right]\delta t + S_d \delta t    
\end{split}
\end{equation}
\begin{equation}
\begin{split}
    P_g &= \xi_g^{i} - \xi_g^{i+1}  + t_c^{-1} \left[r_{gd} \xi_d^{i+1} - \xi_g^{i+1}\right] \delta t \\
    & - c\rho\kappa_g^{i+1}\left[a_r (T_g^{i+1})^4 - \xi_r^{i+1}\right]\delta t + S_g \delta t    
\end{split}
\end{equation}

where 
\begin{equation}
        \xi_r^{i+1} = \xi_r^i -\frac{\hat{c}}{c}\left[(\xi_d^{i+1} - \xi_d^{i}) + (\xi_g^{i+1} - \xi_g^{i})\right] 
\end{equation}
Using Newton's method, we attempt to find a vector $\vec{\xi}^{i+1,*}$, such that $\vec{P}(\vec{\xi}^{i+1,*}) = \vec{0}$ to within some small tolerance $\vec{\alpha}$. The relevant Jacobian matrix $\mathbf{J}(\vec{\xi}^{i+1}) \equiv \partial \vec{P}/\partial \vec{\xi}^{i+1}$ is:
\begin{equation}
    \mathbf{J}(\vec{\xi}_{
\rm sub}^{i+1}) = \begin{bmatrix}
\partial P_d/\partial \xi_d^{i+1} & \partial P_d/\partial \xi_g^{i+1} \\
\partial P_g/\partial \xi_d^{i+1} & \partial P_g/\partial \xi_g^{i+1} \\
\end{bmatrix}
\end{equation}
where
\begin{equation}
    \partial P_d/\partial \xi_d^{i+1} = -(1 + r_{gd} \delta t/t_c) - c\rho\kappa_d^{i+1}f_d \delta t\left[4a_r (T_d^{i+1})^3/\rho f_d c_d + \hat{c}/c\right]
\end{equation}
\begin{equation}
    \partial P_d/\partial \xi_g^{i+1} = +\delta t/t_c - \hat{c}\rho\kappa_d^{i+1} f_d \delta t
\end{equation}
\begin{equation}
    \partial P_g/\partial \xi_d^{i+1} = +r_{gd} \delta t/t_c - \hat{c}\rho\kappa_g^{i+1} \delta t
\end{equation}
\begin{equation}
    \partial P_g/\partial \xi_g^{i+1} = -(1 + \delta t/t_c) - c\rho\kappa_g^{i+1} \delta t\left[4a_r (T_g^{i+1})^3/\rho c_g + \hat{c}/c\right]
\end{equation}
We omit terms dependent on the derivatives of opacity and cooling time, although (these being, in general, functions of temperature) we update them with each iteration of Newton's method. The update goes as:
\begin{equation}
    \mathbf{J}(\vec{\xi}^{i+1}_{n})\left[\vec{\xi}^{i+1}_{n+1} - \vec{\xi}^{i+1}_{n}\right] = -\vec{P}_n
\end{equation}
where $n$ is an iteration index, in contrast to the timestep index $i$.


\section{Self-shadowing instability with long coupling times}\label{sec:ssi_appendix}

Our investigation in Section \ref{sec:ssi} found that, just as in the two-temperature case, our "realistic" three-temperature disk with self-consistently evaluated dust-gas coupling times cannot sustain the self-shadowing instability (SSI). This result, however, leaves open the question of whether there is any circumstance under which three-temperature dynamics could favor the survival of the SSI. To study this possibility, we ran a simulation with dust-gas coupling times 100$\times$ larger than those in the realistic run in Section \ref{sec:ssi}. \revbf{T}o facilitate comparison with our previous results, we hold all other simulation parameters---including, crucially, the \revbf{dust} opacity prescription and dust-to-gas ratio---fixed. \revbf{Formally, solving for $t_s$ in Equation \ref{eq:tcool}, this would imply a Stokes number ${\rm St} \approx 0.06$ for grains at the $\tau_r = 1$ surface, which can be expected to experience vertical drift. However, we emphasize that this $100\times$-cooling setup is artificial, and is designed to approximate and long relaxation times that arise in disk upper atmospheres in detailed calculations of} grain growth, \revbf{settling}, and depletion \revbf{\citep[see e.g.,][for one such scenario]{Bae21}. The dust populations arising from these calculations are, by construction, in steady state, and do not experience net vertical drift.}

\begin{figure}
    \centering
    \includegraphics[width=0.999\linewidth]{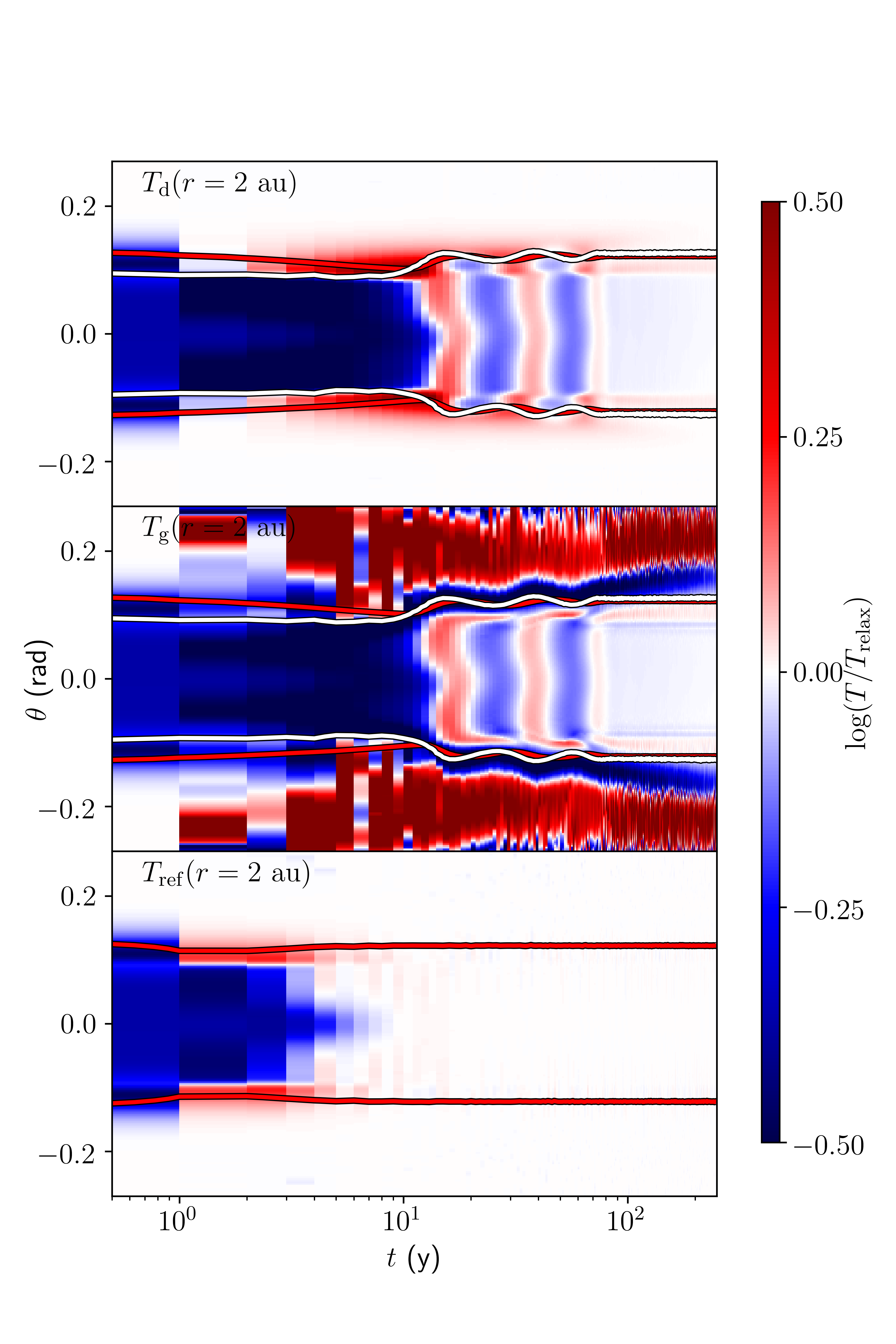}
    \caption{A plot similar to our Figure \ref{fig:ssi_phenomenology}, but with $T_d$ and $T_g$ taken from our simulation with cooling times artificially raised by a factor of 100. There are clear vertical patterns in the dust and gas temperature profiles, corresponding to inward-traveling perturbations at the $\tau_r = 1$ surface characteristic of the SSI.}
    \label{fig:ssi_phenomenology_longcool}
\end{figure}

\begin{figure}
    \centering
    \includegraphics[width=0.999\linewidth]{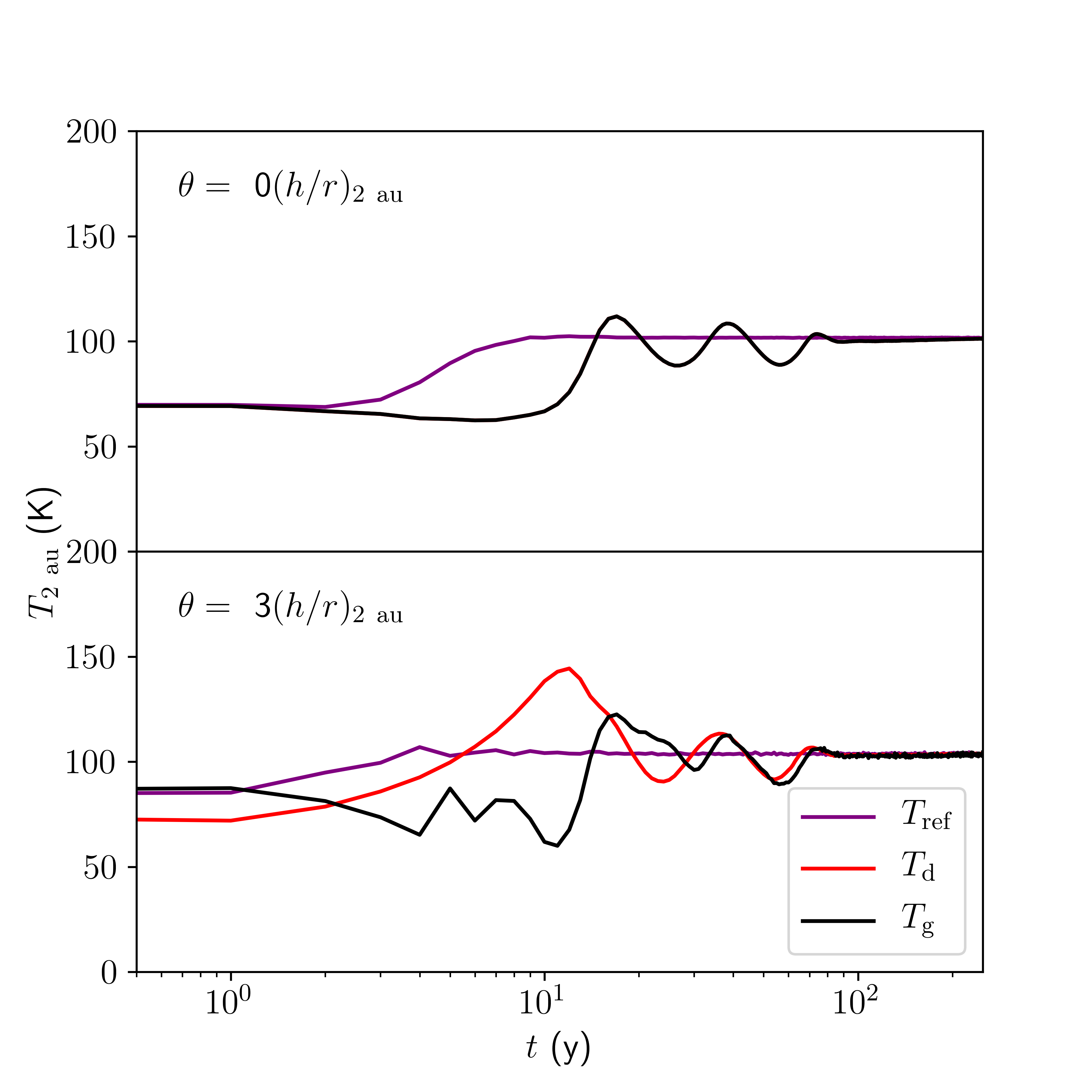}
    \caption{Quantitative time evolution of $T_d$ and $T_g$ from the $100\times$-realistic simulation, and $T_{\rm ref}$ in the reference simulation, at $r = 2$ \ au, both in the midplane ($\theta = 0$, \textit{above}) and 3 scale heights above it ($\theta \approx 0.088$, \textit{below}). As qualitatively shown in Figure \ref{fig:ssi_phenomenology_longcool}, $T_{\rm ref}$ reaches a smooth equilibrium, whereas  $T_d$ and $T_g$ exhibit damped oscillations due to the passage of self-shadowing waves. See text for more details.}
    \label{fig:ssi_quantitative_longcool}
\end{figure}

In Figure \ref{fig:ssi_phenomenology_longcool}, we plot the time evolution of $T_{\rm d}$, $T_{\rm g}$, and $T_{\rm ref}$---normalized by the final, relaxed reference temperature $T_{\rm relax}$---in a fiducial disk column at $r = 2 {\rm \ au}$. In the two-temperature reference simulation, the initialized self-shadowing perturbations rapidly decay (as already evident from Figure \ref{fig:test_self_shadow}), but in the three-temperature 100$\times$ realistic simulation, they survive for some time, as evidenced by the vertical bands in $T_d$ and $T_g$ between $t = 10^1-10^2$ y.

At the leading edge of a self-shadowing scale-height perturbation, the flaring angle of the $\tau_r = 1$ is high and intercepts more direct stellar irradiation. Although $T_d$ reflects this change immediately, the fact that $\beta \gg 1$ in this disk region means that the response of $T_g$---and the consequent vertical expansion of the gas column to seek hydrostatic equilibrium---is significantly slower. As the gas column puffs up, it creates a new leading edge immediately to its interior; when this edge puffs up in turn, it blocks stellar irradiation from reaching the original column, causing it to cool and shrink. In this fashion, self-shadowing waves can propagate inward towards the star purely through vertical motions of the fluid elements. This mechanism would not operate when dust and gas are well-coupled; in that regime, both components would have a common temperature, and would respond to surface perturbations via layered radiative diffusion. The long timescale for this diffusion---given very roughly by Equation \ref{eq:classical_rad_cool})---means that surface perturbations would decay before having an effect deep in the disk \citep{MelonFuksman2022}.

For a more quantitative view, we present in Figure \ref{fig:ssi_quantitative_longcool} the time evolution of $T_d$, $T_g$, and $T_{\rm ref}$ at $r = 2$ au and $\theta = \pi/2 + \{0.0, 0.088\}$ (the latter $\theta$ corresponding to 3 scale heights above the midplane). At both altitudes, $T_{\rm ref}$ relaxes almost immediately to its hydrostatic equilibrium value, where it remains for the rest of the simulation. The evolution of $T_d$ and $T_g$ is more complex. High in the disk, the dust temperature rises interior to self-shadowing bumps, and drops exterior to them; with some delay due to gas cooling parameters $\beta \approx 1$, $T_d$ relaxes toward $T_g$ and the interior and exterior gas columns adapt to the new hydrostatic equilibrium. The net result, as mentioned earlier and in line with typical expectations for the SSI, is the apparent inward propagation of self-shadowing bumps. The changes these bumps make to the disk's illumination profile impact the midplane temperature, albeit with some delay and attenuation due to radiative diffusion.

\begin{figure*}
    \centering
    \includegraphics[width=0.99\textwidth]{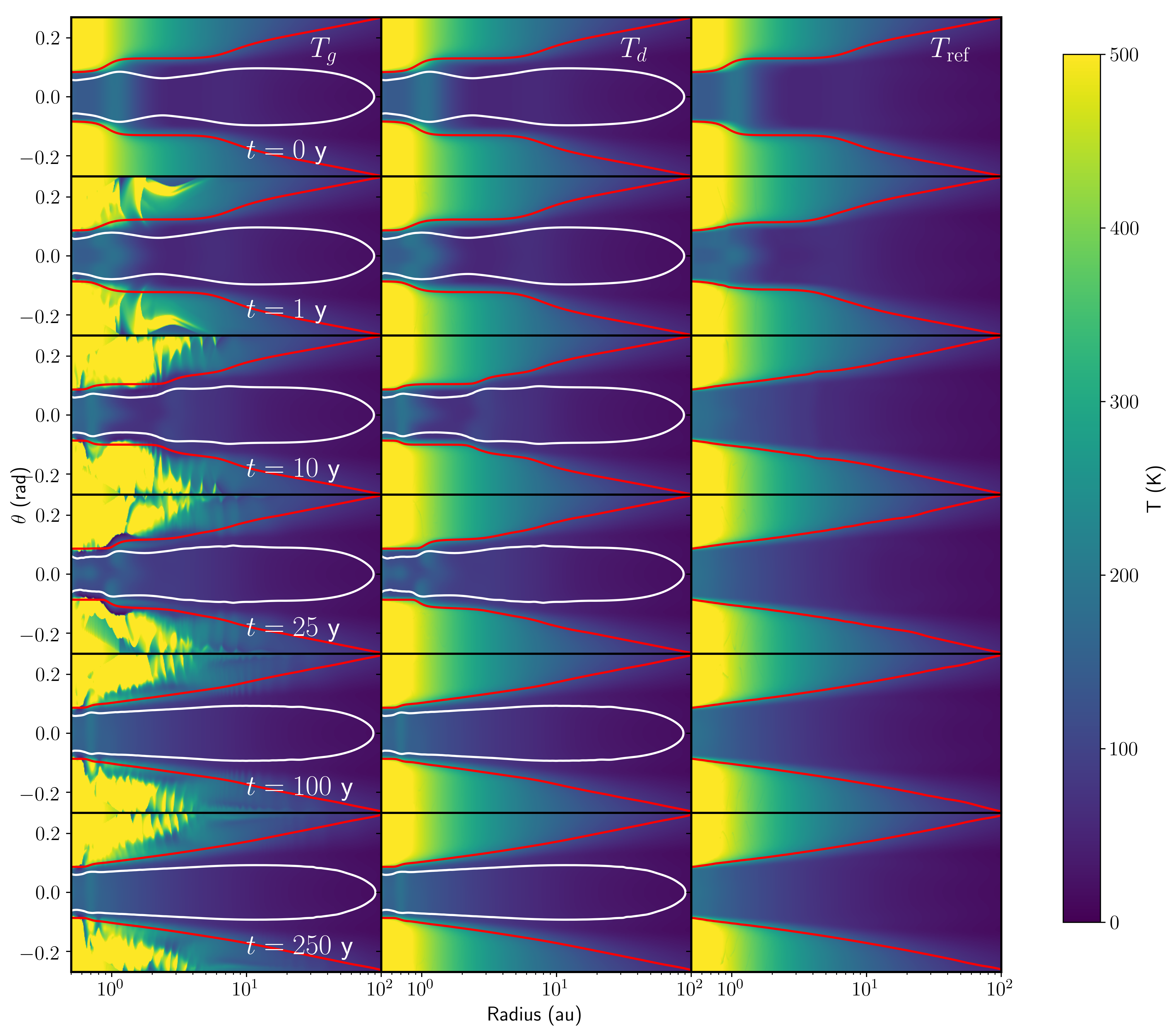}
    \caption{Evolution of the SSI with a dust-gas coupling time 100$\times$ longer than in the self-consistent setup presented in \ref{sec:ssi}. An inward-propagating SSI bump is clearly visible. At $t = 250 {\rm \ y}$, the disk has largely relaxed to the same configuration as in the reference simulation. Differences do persist, however, outside the $\beta = 1$ surface and near the inner boundary.}
    \label{fig:test_self_shadow_longcool}
\end{figure*}

In Figure \ref{fig:test_self_shadow_longcool}, we present snapshots of the full long-coupling disk at selected times. Puffed-up disk columns heated above the background temperature---a hallmark signature of self-shadowing waves---are clearly visible here, and are captured n Figures \ref{fig:ssi_phenomenology_longcool} and \ref{fig:ssi_quantitative_longcool} as they travel inward through $r = 2$ au. By $t = 250$ y, however, the long-cooling disk has largely relaxed to the same smooth equilibrium as the reference disk. Differences persist in the very slowly-cooling disk atmosphere, as well as in the $r \lesssim 1$ au region in the immediate vicinity of our fixed inner radial boundary.

\end{appendix}
\end{document}